\documentclass[a4paper,11pt]{article}
\pdfoutput=1 

\usepackage{jheppub} 

\newcommand{\PR}{{\em Phys. Rev. }}
\newcommand{\PRL}{{\em Phys. Rev. Lett. }}
\newcommand{\PL}{{\em Phys. Lett. }}
\newcommand{\NP}{{\em Nucl. Phys. }}
\newcommand{\etal}{{\em et al}}
\newcommand{\AP}{{\text A}^\prime}

\def\AB{\mbox{A$^\prime$-boson}}
\def\V3{\mbox{VEPP--3}}
\def\mb{m_{_{{\text A}^\prime}}}

\graphicspath{ {./figs/} }

\title{Searching for a dark photon: Project of the experiment at \V3}

\author[a]{B.~Wojtsekhowski}
\author[b]{G.N.~Baranov}
\author[b]{M.F.~Blinov}
\author[b,c]{E.B.~Levichev}
\author[b]{S.I.~Mishnev}
\author[b]{D.M.~Nikolenko}
\author[b]{I.A.~Rachek}
\author[b,d]{Yu.V.~Shestakov}
\author[b,d]{Yu.A.~Tikhonov}
\author[b,d]{D.K.~Toporkov}

\author[e] {J.P.~Alexander}

\author[f] {M.Battaglieri}
\author[f] {A.Celentano}
\author[f] {R.De Vita}
\author[f] {L.Marsicano}

\author[g] {M.Bond\`i}
\author[g] {M. De Napoli}
\author[g] {A.Italiano} 
\author[g] {E.Leonora}
\author[g] {N.Randazzo}

\affiliation[a]{Thomas Jefferson National Accelerator Facility\\ Newport News, VA 23606, USA }
\affiliation[b]{Budker Institute of Nuclear Physics\\ 630090 Novosibirsk, Russia}
\affiliation[c]{Novosibirsk State Technical University\\ 630073 Novosibirsk, Russia}
\affiliation[d]{Novosibirsk State University\\ 630090 Novosibirsk, Russia}
\affiliation[e]{Cornell University\\ Ithaca, New York, 14853, USA}
\affiliation[f]{Istituto Nazionale di Fisica Nazionale - Sezione di Genova\\ Genova, Italy}
\affiliation[g]{Istituto Nazionale di Fisica Nazionale - Sezione di Catania\\ Catania,  Italy}

\emailAdd{bogdanw@jlab.org}
\emailAdd{D.M.Nikolenko@inp.nsk.su}
\emailAdd{rachek@inp.nsk.su}

\abstract{
We propose an experiment to search for a new gauge boson $\AP$ in $e^+e^-$ annihilation
by means of a positron beam incident on a gas hydrogen target internal
to the bypass at the \V3 storage ring.  
The search method is based on a missing mass spectrum in the reaction
$e^+e^-\rightarrow \gamma\AP$.
It allows observation of the $\AP$~signal independently of its decay modes and life time.
The projected result of this experiment corresponds to an upper limit on the square
of the coupling constant $\varepsilon^2=3\cdot 10^{-8}$ with a signal-to-noise ratio of two
to one at an $\AP$~mass of 5-20 MeV.
}


\begin{document}
\maketitle
\flushbottom

\section{Introduction}
The search for an experimental signature of physics beyond the Standard Model
is a major effort of modern particle physics, see e.g.~\cite{pdg4}.
Most of the search activity is focused on possible heavy particles with 
a mass scale of 1~TeV and above.  
At the same time, as was suggested by P.~Fayet~\cite{fa80,fa90}, there could be  
extra $\text{U(1)}$ symmetry, which requires a new gauge boson, U, also called the \AB.  
The boson could be light and weakly interacting with known particles through kinetic mixing
with the ordinary photon \cite{holdom}.
Most constraints for the light \AB~parameters were obtained from
electron and muon anomalous magnetic moments $(g-2)$ and 
particle decay modes \cite{fa06,fa07,fa08,po09}.

Renewed interest in a search for the new gauge boson has been seen recently as 
such a boson may provide an explanation for various astrophysics phenomena, 
accumulated during the last decade,  which are related to dark matter %
\cite{fa06,fa07,arkani,est09}.  
The possible connection between the \AB~and dark matter in view of the
observed slow positron abundance has been investigated for several years 
and is often referred to as MeV dark matter 
\cite{bo04,fa04,bofa,bo06}.  
The theory of dark matter proposed by N.~Arkani-Hamed and
collaborators~\cite{arkani}, which provided interpretation of a number
of key astrophysical observations, 
sparked additional interest in an \AB~search in the mass range below 1~GeV.  

Several methods have been used in the search for the \AB~signal,
considering ``invisible'' decay modes of the \AB.  
The first method uses precise experimental data on exotic
decay modes of elementary particles, e.g. 
\mbox{$\pi^\circ \rightarrow invisible + \gamma$},
for the calculation of the upper limit on the \AB~coupling constant $\varepsilon$
to the specific flavour.  
These upper limits for decay of the $J/\Psi$ and $\Upsilon$ to a photon
plus invisible particles were obtained experimentally
by means of the ``missing particle'' approach, where 
a missing particle in the event type $e^+e^- \rightarrow \gamma X$
leads to a yield of events with a large energy photon detected at 
a large angle with respect to the direction of the positron 
and electron beams.  
From the yield of such events the coupling constant could be determined
for a wide range of mass of the hypothetical \AB.  
A recent experiment~\cite{babar17} using 53 fb$^{-1}$ of $e^+e^-$ collision data collected with CM energies near 
the $\Upsilon(2S)$, $\Upsilon(3S)$ and $\Upsilon(4S)$ resonances provided the best data for $\Upsilon$ decay to $\AP + \gamma$ and a 
limit on the coupling of the \AB~to the $b$-quark.
In the mass region below 100~MeV the obtained limit for 
$B(\Upsilon(2S,3S,4S) \to \gamma \AP) \times B(\AP \to \text{invisible})$ is $1 \cdot 10^{-6}$;
an additional hypothesis of coupling constants' universality is required 
to get a bound on $\varepsilon$, so direct measurement 
of the coupling to an electron is of large interest.  
Currently, the upper limit on the vector coupling 
obtained from the discrepancy between the calculated electron anomalous magnetic moment
and the measured one is $\varepsilon < 1.0 \cdot 10^{-4} \, \times \mb[\text{MeV}]$ \cite{fa07,po09}.

One more approach to searching for the \AB~decaying to invisible states is the missing-energy method, 
used by the NA64 collaboration at CERN SPS \cite{na64}.
Such a method provides a very high sensitivity in a wide range of \AB~mass but there is a potential 
problem due to use of a veto approach which suppresses observation of a semi-invisible decay.
Similar decay modes were considered recently in Ref.~\cite{ch18}.
In the missing-energy  method the parameters  $\mb$ and $\varepsilon$ can not be separately extracted.  
This limitation is overcome in the LDMX experiment \cite{LDMX}, which performs tracking on the recoiling electron.

A direct measurement of $\varepsilon$ and $\mb$ could be made by detecting 
the decay of the \AB~to an electron-positron pair and reconstructing 
the $e^+e^-$ invariant mass.  
It requires a significant branching of \AB~decay to the $e^+e^-$ pair.
A complication of this method is the high level of 
electromagnetic background in the mass spectrum of $e^+e^-$, 
so such a measurement requires large statistics.
Recently the data sets accumulated in collider experiments  
have been used for such an analysis \cite{bo06,babar,kloe}.

Electron fixed-target experiments, where a new boson can be produced from radiation off
an electron beam incident on an external target, are now widely
discussed \cite{ap2,ap3,ap4,apex1}.  
The first significant experimental results on upper limits for a new boson coupling to an electron
in the sub-GeV mass range have been reported \cite{mainz-T,apres}.  
The APEX experiment in JLab Hall A \cite{apex2} will probe couplings 
$\varepsilon^2 > 10^{-7}$ and masses $\mb\sim 50 - 550$~MeV.  
The result of the test run, with only 1/200 of the data of the full APEX experiment,
has already demonstrated the feasibility of such an approach~\cite{apres}.
A full scan in mass range $\mb\sim 40 - 300$~MeV was performed at MAMI \cite{mainz}, which
put an upper limit on the coupling $\varepsilon^2 < 0.8 \times 10^{-6}$.
Other electron fixed-target experiments are planned: \  at Jefferson Lab,
including the Heavy Photon Search (HPS)~\cite{HPS} and DarkLight~\cite{darklight},
and at the MESA facility at Mainz~\cite{mainz-M}. 

A high sensitivity \AB~search could be performed with a low energy 
$e^+e^-$ collider \cite{est09}, where several search techniques could be used:
\begin{itemize}
\item The invisible particle method.
\item Invariant mass of the final $e^+e^-$ pair.
\item Missing mass with single-arm photon detection. 
\end{itemize}

To search for the \AB~with a mass of 10-20~MeV 
the center-of-mass energy of $e^+e^-$, $E_{cm}$, should be low.
The production cross section is proportional to $1/E_{cm}^2$, so for 
low $E_{cm}$, even a modest luminosity would be sufficient for 
a precision measurement.

The design of a new collider with high luminosity in the mass range of interest 
for the dark photon search (below 1 GeV) could use the sliding beam configuration 
where two high energy beams collide at a small angle between themselves~\cite{BW_DF}.
In such an unusual configuration the invariant mass could be adjusted by a change of 
the angle but the luminosity would be high thanks to the compact beam parameters.
Another interesting option for a high luminosity collider at modest energy could be 
with an energy recovery linac which provides an electron beam of high intensity
with very small emittance. 
When such a beam collides (head-to-head) with a high energy positron beam (e.g. 5 GeV) 
the invariant mass could be in the range of 1-3 GeV and its luminosity will likely match
the luminosity of a standard 5 GeV x 5 GeV collider.

However, for the low energy region (below 100 MeV) no colliding ($e^-e^+$)--beam facility 
exists or is planned to be constructed.
Still, a similar operation can be achieved if an available positron beam of
a few hundred MeV energy is incident on a fixed target \cite{wo06, ref:concept}. 
The \V3 electron/positron storage ring at the Budker Institute in Novosibirsk \cite{vepp} 
with its internal target facility and high-intensity positron beam injection complex  
is uniquely suited for such measurements.
\vskip+\baselineskip

\emph{
We propose to perform a search for the \AB~in a mass range 
$\mb=5-20$~MeV 
using a 500~MeV positron beam incident on an internal hydrogen target, providing a luminosity of  
$10^{33}$~cm$^{-2}$s$^{-1}$, 
by detecting $\gamma$-quanta from the process $e^+e^-\rightarrow \gamma \AP$ 
in an energy range  $E_\gamma=50-450$~MeV and angular range 
$\theta_\gamma^{CM}=90^\circ\pm 30^\circ$ ($\theta_\gamma^{Lab}=1.5^\circ - 4.5^\circ$).
}

\section{The concept}
In the proposed experiment we would like to explore the technique of the missing mass
measurement approach with a single-arm photon detector. 
The concept of the method is partly described in~\cite{ref:concept}.
A positron beam in a storage ring with an energy $E_+$ of a few hundred MeV and an internal
hydrogen gas target make up an ``$e^+e^-$ collider''. In such a collider it is possible
to search for a light \AB~with a mass of up to $\mb[\text{MeV}]\sim \sqrt{E_+[\text{MeV}]}$.
Unlike all other experiments with a fixed target which are based on the detecting of $e^-e^+$ pairs
from \AB~decay, 
\textbf{\textit{in the proposed experiment  no special assumptions about decay modes 
of the \AB~are required.}} 
In this proposal we consider a medium luminosity ($\sim 10^{33}$~cm$^{-2}$s$^{-1}$) 
measurement and a combination of on-line and off-line veto on the bremsstrahlung 
and multi-photon background processes.

\begin{figure}
  \centerline{\includegraphics[width=0.7\linewidth]{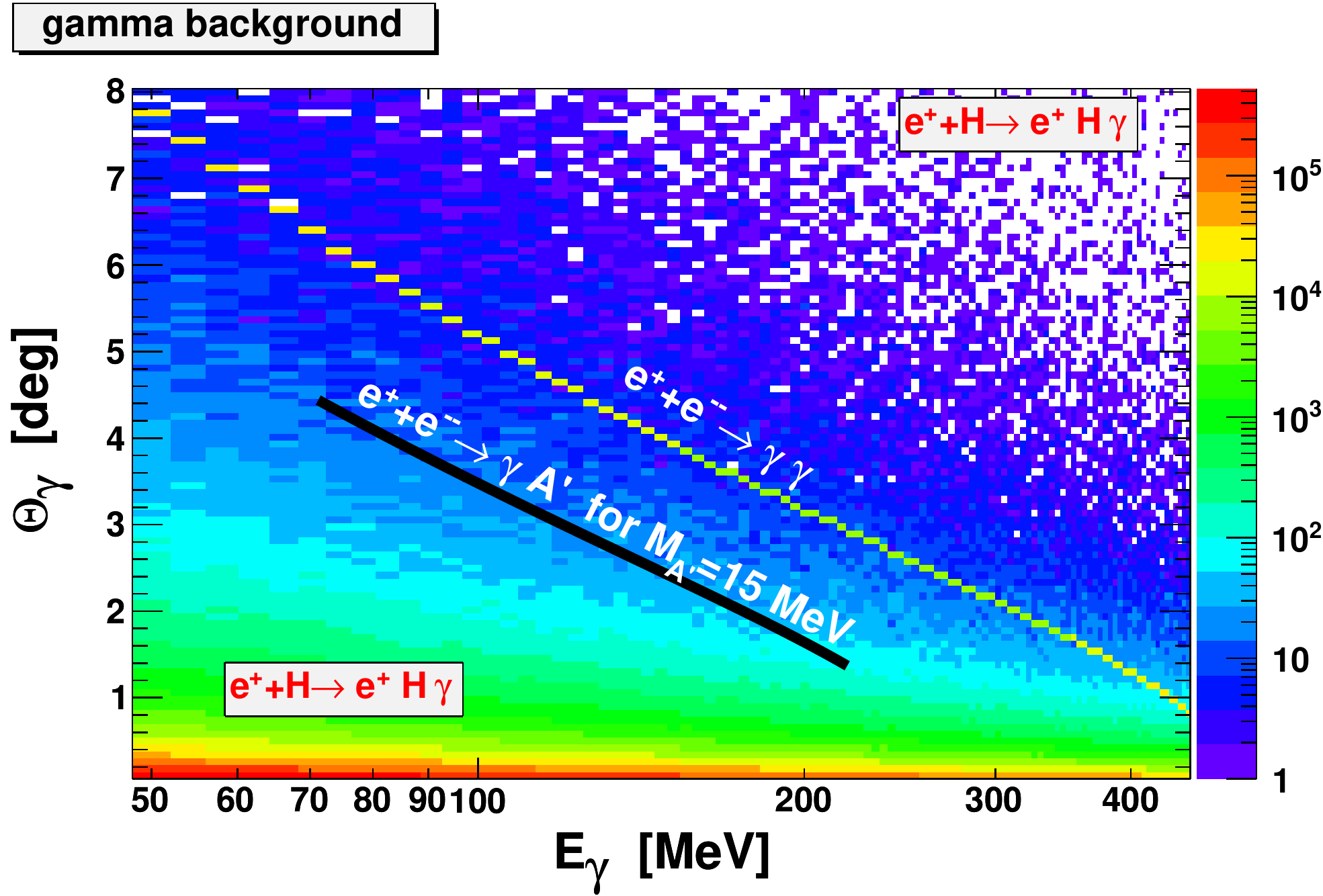}}
\caption[]{\label{fig:2d}
Two-dimensional distribution of the photon events in the scattering angle and the 
photon energy for a 500 MeV positron beam incident on a hydrogen gas target. 
The black band shows the location of \AB~events of 15 MeV mass. 
}
\end{figure}

In the process $e^+e^-\rightarrow \AP\gamma$ a measurement of the photon energy and its 
angle allows a reconstruction of the missing mass spectrum and a search for a peak
corresponding to the \AB. In such a spectrum the dominant signal corresponds to the 
annihilation reaction $e^+e^-\rightarrow \gamma\gamma$. The signal for the \AB~will be
shifted to the area of the continuum (see the illustration in Fig.~\ref{fig:2d}).
The continuum part of the event distribution is dominated by photons emitted 
in the process of positron scattering from an electron or a proton in the target 
(bremsstrahlung) and by photons from the three-photon annihilation process. 
Contributions of other reactions, e.g. 
$\gamma^\star p \rightarrow p \pi^0 \rightarrow p \gamma\gamma$,
are at least three orders of magnitude smaller than that of positron bremsstrahlung.

A key property of the proposed experimental setup is the ability to suppress the QED background 
significantly, both on-line and off-line, thus improving the sensitivity of the search. 

\section{The kinematics and cross sections for the signal and background}
\begin{figure}
        \includegraphics[width=0.45\linewidth]{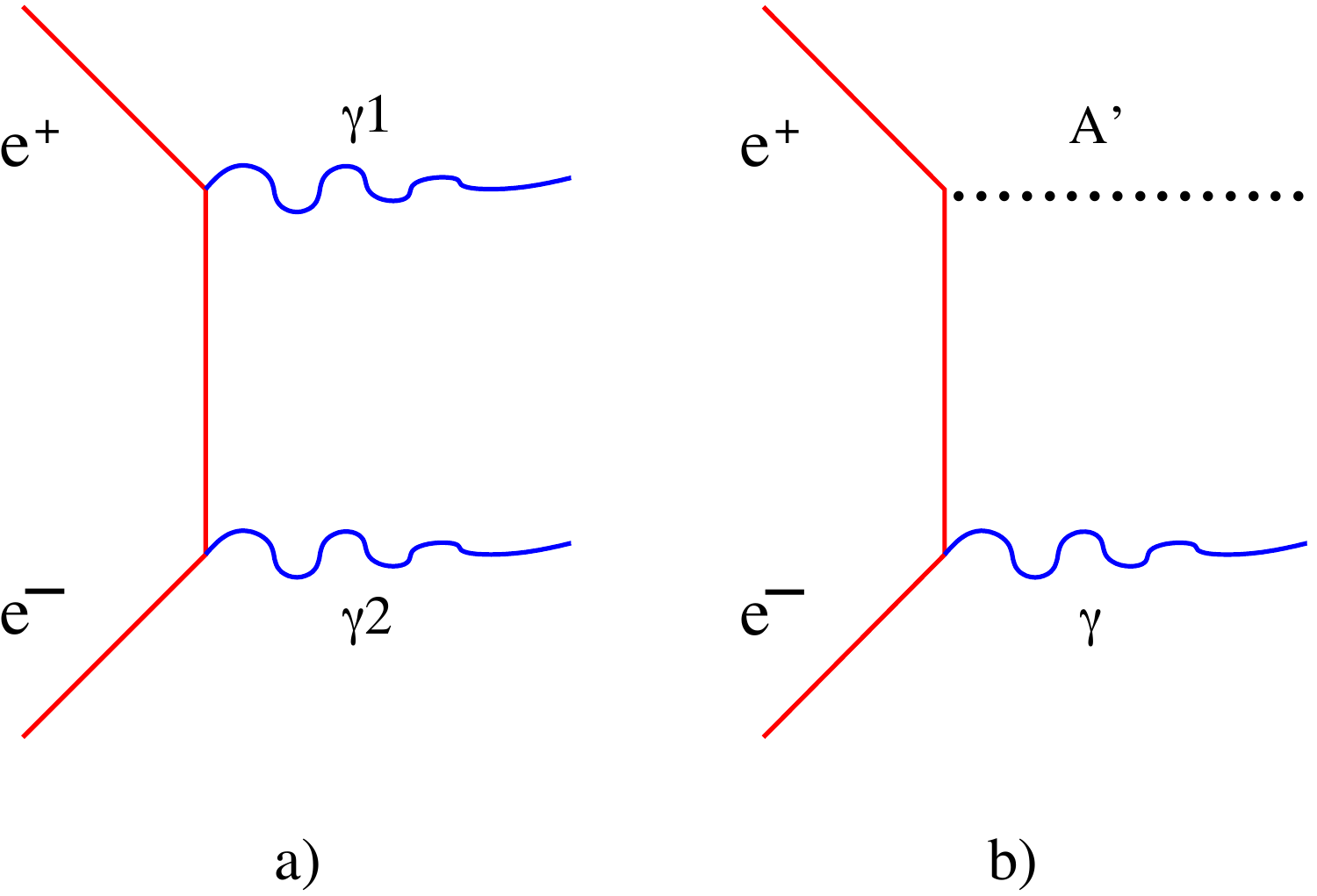}
\hfill
     \includegraphics[width=0.4\linewidth]{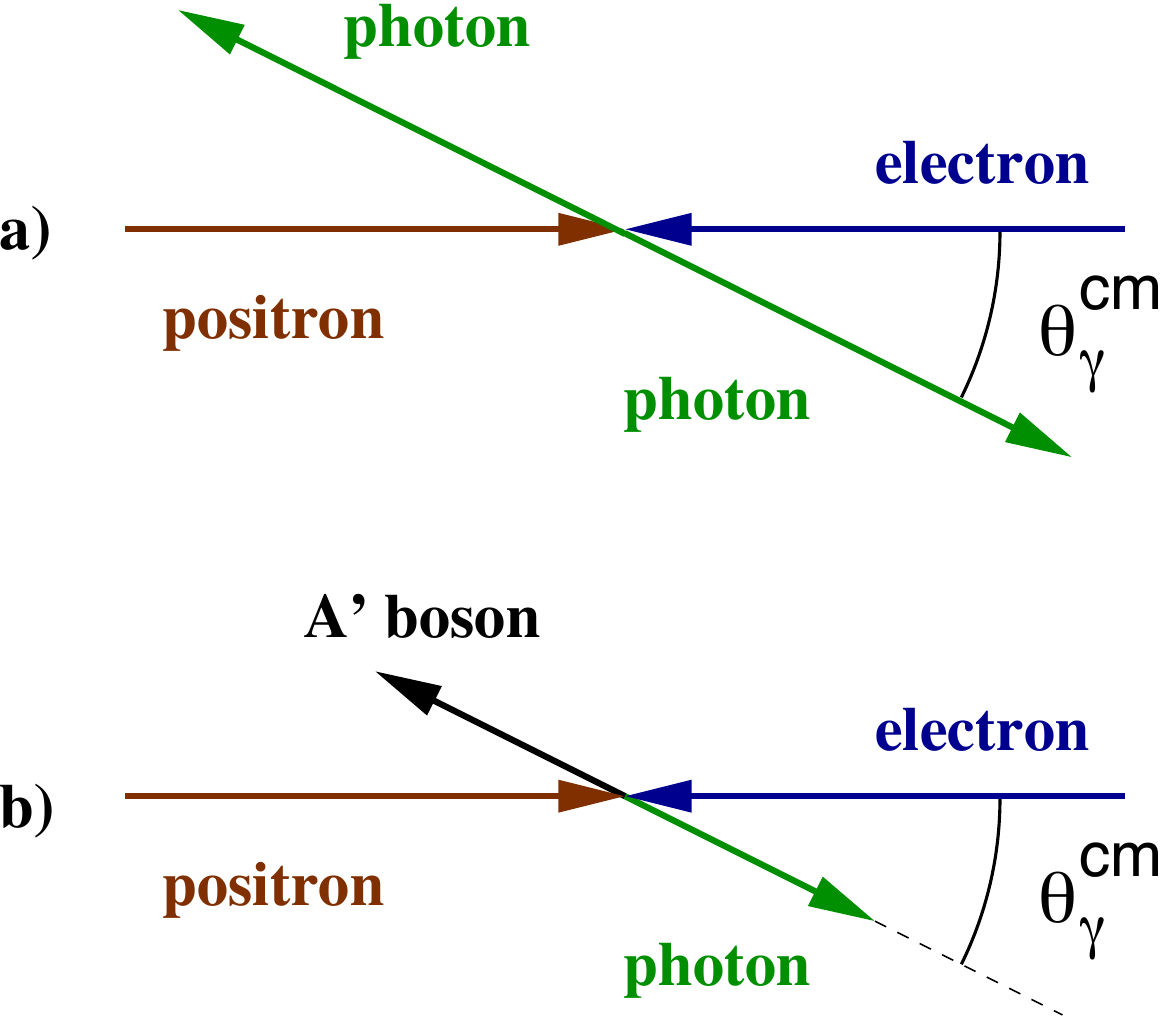}
\caption{
\label{fig:two-body}
The diagrams (on the left) and the kinematics (on the right) of a) two-photon annihilation: 
$e^+e^- \to \gamma + \gamma$, and b) \AB\ production:  $e^+e^- \to \AP + \gamma$.
}
\end{figure}

Two-photon annihilation is the dominant process of high-energy photon 
production in $e^+e^-$ collisions at a cms energy of a few tens of MeV.
Two reactions, depicted in the left panel of Fig.~\ref{fig:two-body}, 
are two-photon annihilation and the production of an exotic \AB. 
The kinematics for the two-body final state is shown in the right panel
of Fig.~\ref{fig:two-body}.
The energy in the center of mass system 
$\sqrt{s} \,=\,  \sqrt{2m^2 \,+\, 2E_+ m} $,
where $m$ is the electron mass and $E_+$ the positron energy,  
and the emission angle of the final photon $\theta_\gamma$ 
with respect to the direction of the positron beam
defines the value of the photon energy $E_\gamma$.
In the case of two-photon production: 
$E_{\gamma (\gamma\gamma)}^{lab} \approx E_+ (1 \,-\, \cos\theta_\gamma^{cm})/2$.
In the case of \AB~production: 
$E_{\gamma(A^\prime\gamma)}^{lab} \,=\, E_{\gamma (\gamma\gamma)}^{lab}
\cdot (1 \,-\, M_{A^\prime}^2/s)$.
The kinematic boost from the center of mass system to the lab leads to a larger photon
energy in the forward direction, which helps the measurement of 
the photon energy.
The large variation of the photon energy with the photon angle 
in the lab system provides an important handle on the systematics. 

\begin{figure}
\includegraphics[width=0.48\textwidth]{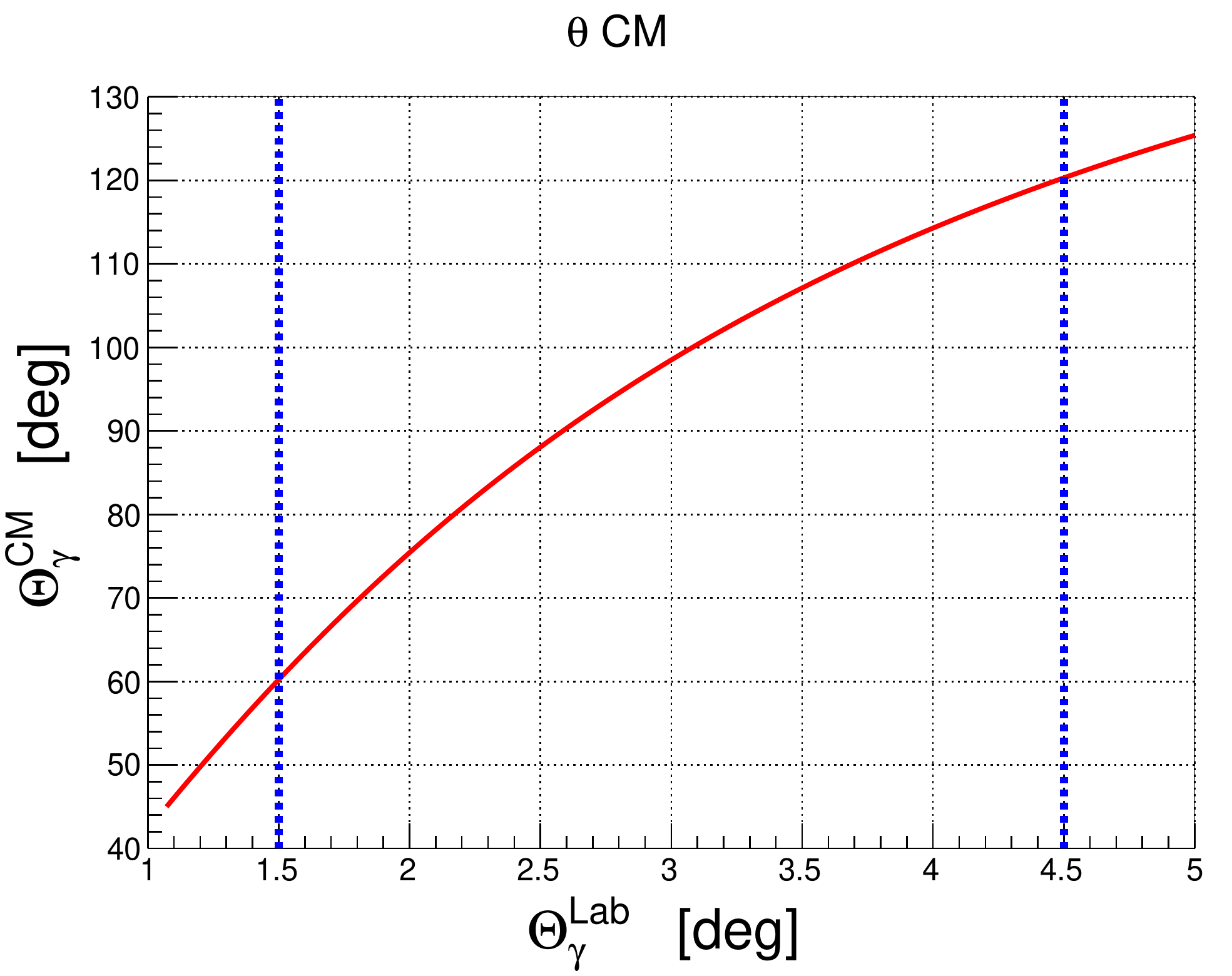}\hfill
\includegraphics[width=0.48\textwidth]{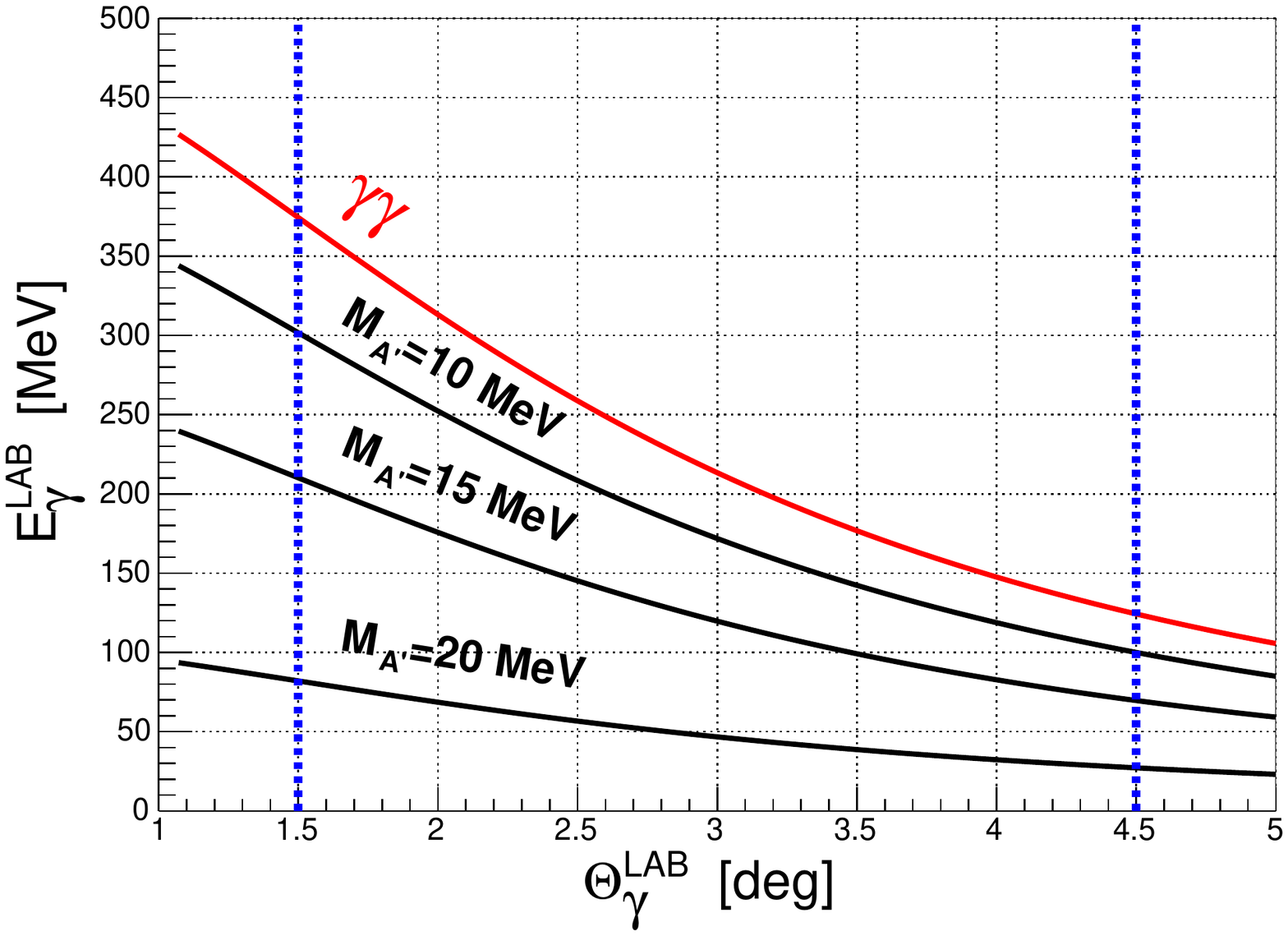}
\caption[]{\label{fig:kinplots}
Kinematic correlations for positron-electron annihilation at E$_+=500$~MeV.
Left panel: photon polar angle in CM frame vs. that in Lab frame for two--photon 
annihilation. Right panel: photon energy vs. its polar angle for 
$e^+e^-\to \gamma\gamma$ and for $e^+e^-\to \gamma \AP$.
Dotted vertical lines indicate the range covered in the proposed measurements.
}
\end{figure}

Figure \ref{fig:kinplots} shows some correlations between kinematic variables for
the proposed setup at \V3.

The energy spectrum of the photons from the two-photon annihilation process in the lab frame 
is expressed by~\cite{ref:heitler}:
\begin{subequations}
\begin{equation}
\label{twogamma}
\frac{d\sigma}{dy} = \frac{\pi r^2_e}{2\gamma_+-2}
\left\{
  \frac{1}{y}
\left[
1 -  y - \frac{2\gamma_+y -1}{y(\gamma_++1)^2}
\right]
+ \frac{1}{1-y}
\left[
y - \frac{2\gamma_+(1-y)-1}
{(1-y)(\gamma_++1)^2}\right]
\right\},
\end{equation} 
where $y = E_\gamma^{lab}/(E_++m)$, with 
${y_{min}} =1/2\left[1-\sqrt{(\gamma_+-1)/(\gamma_++1)}\right]$ 
and ${y_{max}} = 1 - y_{min}$. 
In the case of high-energy positrons ($\gamma_+\gg1$) the expression can be simplified to:
\begin{equation}
\label{twogamma_simple}
\frac{d\sigma}{dy} \approx \frac{\pi r^2_e}{2\gamma_+}
\left[
  \frac{1-y}{y} + \frac{y}{1-y}
\right].
\end{equation}
\end{subequations}

The differential cross section for the process of \AB\ production in the limit of $\gamma_+\gg 1$ 
can be derived for the lab frame from \cite[Eq.55]{fa07}:
%
\begin{equation}
\label{csect}
\frac{d\sigma}{dy} \approx \varepsilon^2 \cdot 
\frac{\pi r^2_e}{y\gamma_+} \ 
\left[\frac{(1+\mu)^2}{1-(y+\mu)}-2y\right],
\end{equation} 
where $\mu=\mb^2/s$ \ and here the photon energy is limited by $y<(1-\mu)$.%
\footnote{Note that the $2\gamma$ cross section (\ref{twogamma_simple}) can be derived 
from (\ref{csect}) by setting $\mu=0$, $\varepsilon=1$ and multiplying by 1/2 to account for two identical photons in the final state.}

\begin{figure}
\includegraphics[width=0.48\textwidth]{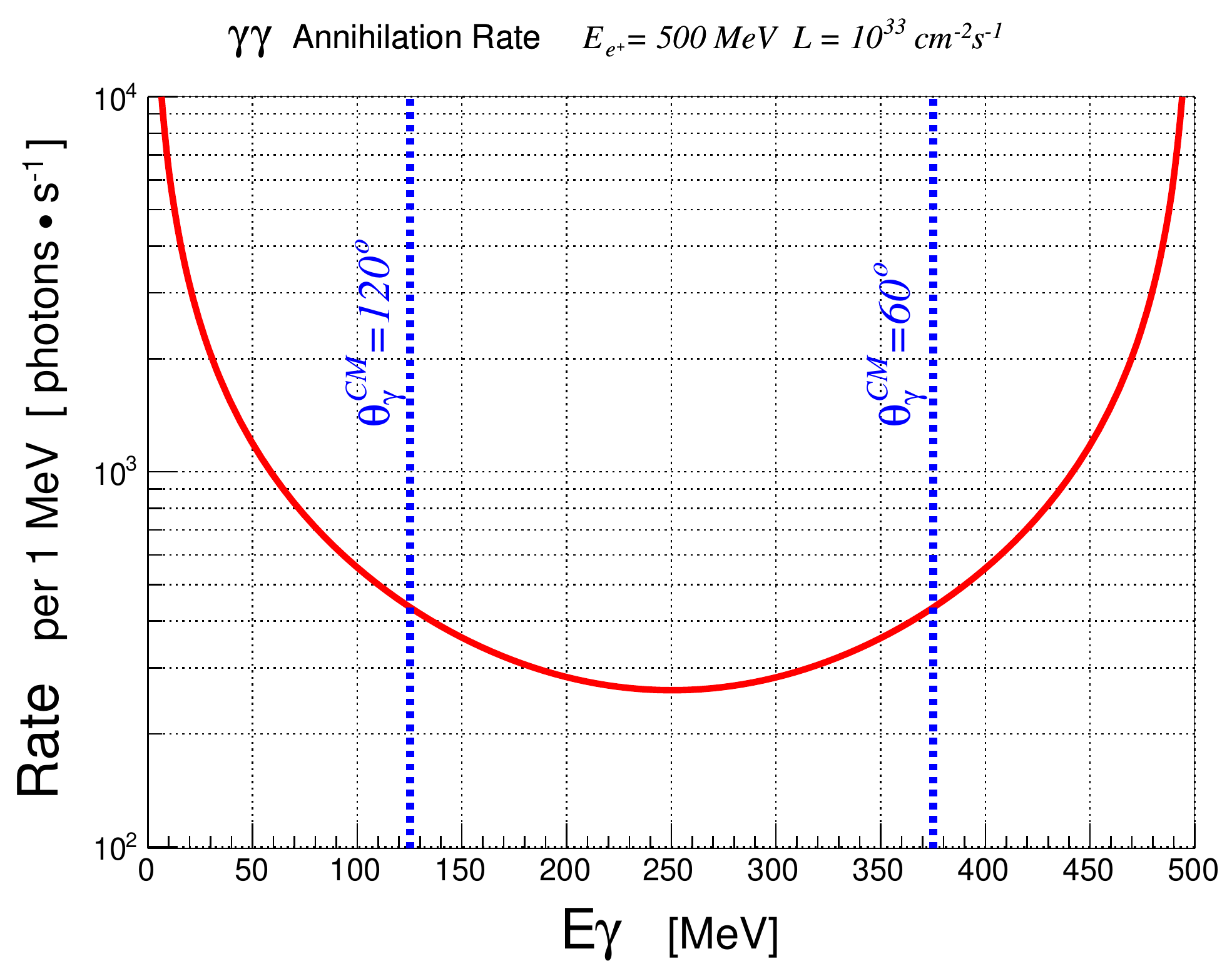}
\hfill
\includegraphics[width=0.48\textwidth]{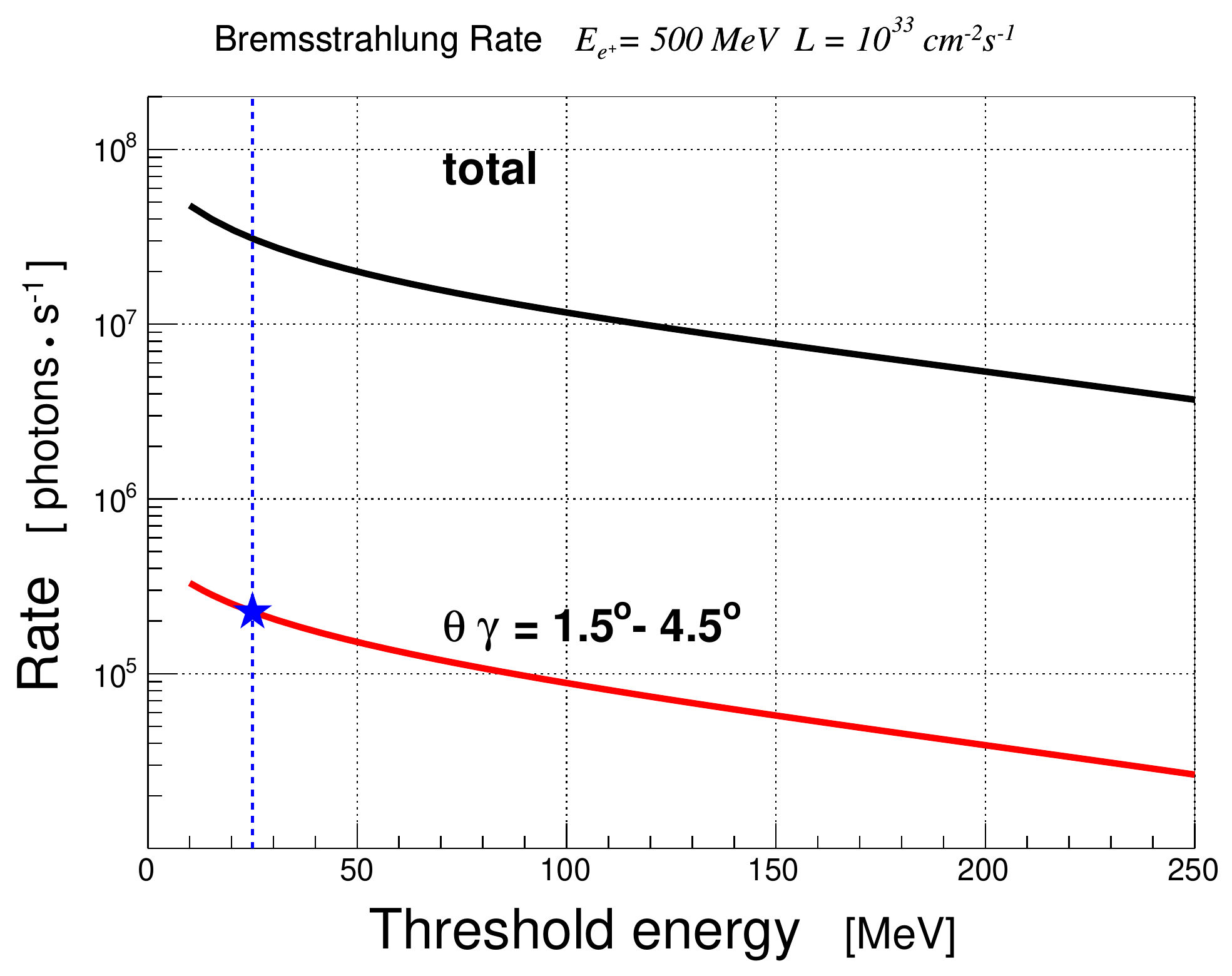}
\caption[]{\label{fig:rates}
Expected background photon rates at beam energy E$_+=500$~MeV and luminosity
$L=10^{33} \, cm^{-2}s^{-1}$. 
Left panel: from the two-photon positron-electron annihilation (Eq.~\ref{twogamma}). 
Dotted vertical lines indicate the range covered in the proposed measurements.
Right panel: from positron bremsstrahlung on hydrogen (Eq.~\ref{eq:tsai}).
At a 25 MeV threshold, the expected rate for the proposed detector configuration is
$2.3\cdot 10^5~s^{-1}$. 
}
\end{figure}

The main physical background process producing a single photon, hitting the photon 
detector, is the positron bremsstrahlung. The differential 
cross section of this reaction in the case of a thin hydrogen target can be evaluated using the  
expression from Ref.\cite{ref:tsai}:
\begin{eqnarray}
\nonumber
\frac{d\sigma_\gamma}{dy d\Omega_\gamma}  =  
\frac{4\alpha r^2_e}{\pi}\frac{\gamma_+^2}{y}
\left\{\frac{2y-2}{\left(1+l\right)^2} 
\right. 
 &+&
\left. 
\frac{12l(1-y)}{\left(1+l\right)^4}
 +\left[\frac{2-2y+y^2}{\left(1+l\right)^2} - \frac{4l(1-y)}{\left(1+l\right)^4}
 \right]
\right. \\
\label{eq:tsai}
& \times &
 \left.
 \left[1+2\ln{\frac{2\gamma_+(1-y)}{y}}-\left(1+\frac{2}{B^2}\right)
 \ln{\left(1+B^2\right)} 
 \right]
\right\}, 
\end{eqnarray}
where $l=\gamma_+^2\theta_\gamma^2$, $B= 4\alpha \gamma_+(1-y)/y(1+l)$. 
The expected rate of photons from these processes is shown in Fig.~\ref{fig:rates} 

However, for the correct reproduction of the $\gamma$-quanta angular distribution at $l>1$, 
one should consider  more accurate formulas for positron-electron elastic (Bhabha) scattering with bremsstrahlung.
We employed a simplified version of the approach described in \cite{CMD}.

One more process which may produce a non-negligible background rate in the considered search is a 3-photon annihilation $e^+e^-\to\gamma\gamma\gamma$. 
With sufficient accuracy it can be evaluated as a radiative correction to the dominating two-photon annihilation process. 
We used the prescription from Ref.\cite{berends} to account for both $\gamma\gamma$ and $\gamma\gamma\gamma$ annihilation channels in a consistent way.

\section{The proposed experimental setup}
\V3 is a booster--ring, operating as an intermediate accelerator/storage ring of electrons 
and positrons for the VEPP-4 collider. The recently commissioned new VEPP-5 electron/positron injection complex 
is able to provide an injection rate of $2\times10^9$ positrons per second at an energy of $E_+\approx 500$~MeV.
In addition, the injection complex includes a damping storage ring at the final stage. 
This allows the implementation of an effective 6-bunch injecting scheme  in which the oldest bunch is replaced by a new one every 10 seconds. 
Each new bunch contains $2\times10^{10}$ $e^+$, corresponding to a VEPP-3 beam current of 13~mA. 
For an internal hydrogen target having a thickness of $10^{16}$ atoms/cm$^2$, the beam lifetime in \V3 is 60 seconds. 
With 6 sequentially refilled bunches this gives an average beam current of $\sim 50$~mA and a luminosity of $\sim 3\times10^{33}$~cm$^{-2}$s$^{-1}$.

Usually the internal target is located in one of the two 12-meter-long straight sections of the \V3 ring. 
In the same straight section there are also two RF cavities, four quadrupoles and one sextupole lens, and elements of beam injection and extraction channels. 
The space available for the internal target equipment is 217~cm long. 
In our earlier proposal \cite{v3arxiv} we described a possible configuration of the experiment based on a chicane magnet installed in this segment of \V3.
However, such a configuration has significant difficulties, both technical and organizational. 
The latter follows from the fact that when a chicane magnet is installed in the ring, 
\V3 becomes completely inaccessible for other working regimes, including operation as a booster for VEPP--4 or as a synchrotron radiation source. 

Therefore, for the proposed experiment we are going to implement a special setup. 
The main element of this setup is a {\bf bypass}, which has to be built along one of the straight sections of the \V3 ring.

\subsection{Bypass}
\label{bypass}
\begin{figure}
  \centerline{\includegraphics[width=\textwidth]{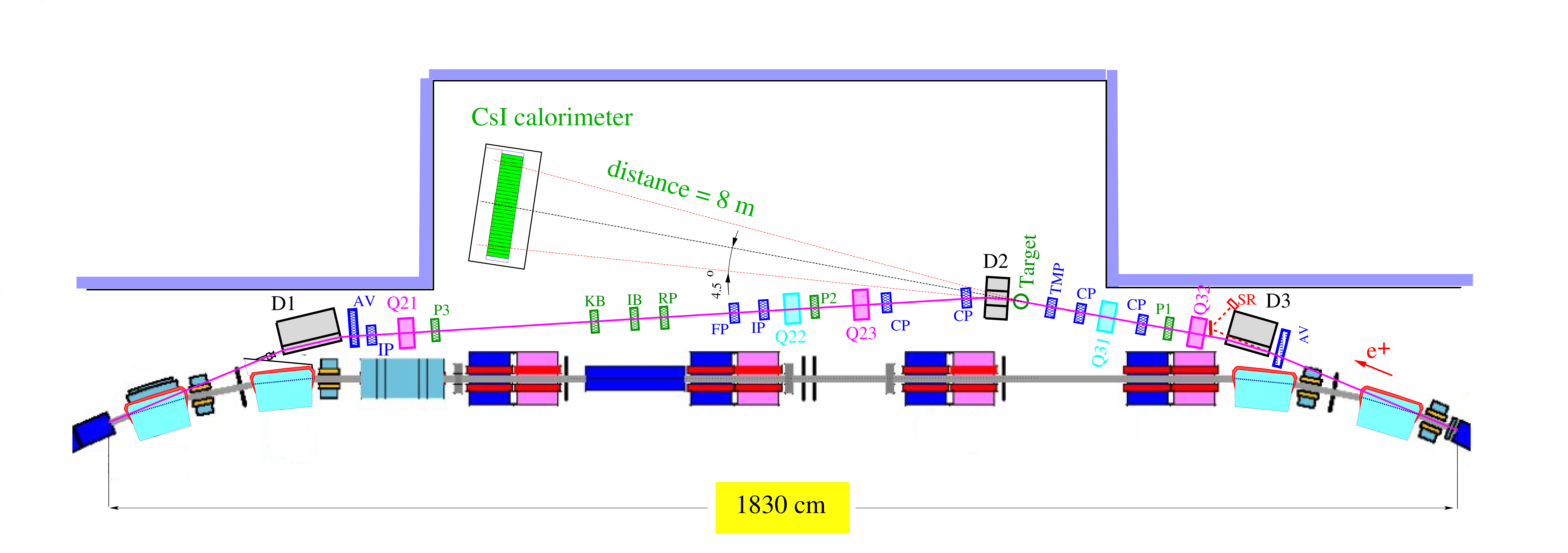}}
\caption[]{\label{fig:bypass}
The layout of the proposed experiment at \V3.
}
\end{figure}

The proposed configuration of the bypass is shown in Figure~\ref{fig:bypass}.
It consists of 
{\it i)} a vacuum chamber with a total length of about 18~m,
{\it ii)} vacuum pumps,
{\it iii)} three dipole magnets with a total rotating angle of $45^\circ$,
{\it iv)} five quadrupole lenses and several correction magnets,
{\it v)} elements of beam diagnostics and
{\it vi)} an internal target section containing a thin-walled storage cell. 

The bypass will occupy the place which several decades ago was already used for another bypass containing the Free Electron Laser. 
The FEL has been dismantled, but the elements providing its operation at VEPP--3  (switching magnets, inlet/outlet vacuum channels) are still there. 
This should simplify significantly the commissioning of the new bypass.

In order to allow the photons of the positron-electron annihilation to pass to
the detector without obstruction, a dipole magnet D2 with a large vertical aperture must be installed immediately after the target. 

Several magnetic elements  can be chosen from the magnets existing in BINP (refurbished or spare ones available from other projects) for no cost. 
However, the D2 magnet has to be constructed from scratch because its 
aperture must be significantly larger than the one normally used in conventional accelerator dipoles. 
 
When VEPP-3 switches to operation with the proposed bypass, the length of its orbit increases by 34 cm. Therefore, the frequency of 
the VEPP-3 RF cavity has to be decreased by 0.45\%. This is within its working range, so no modification of the VEPP-3 RF system is required.
\begin{figure}
\centerline{\includegraphics[width=0.65\linewidth]{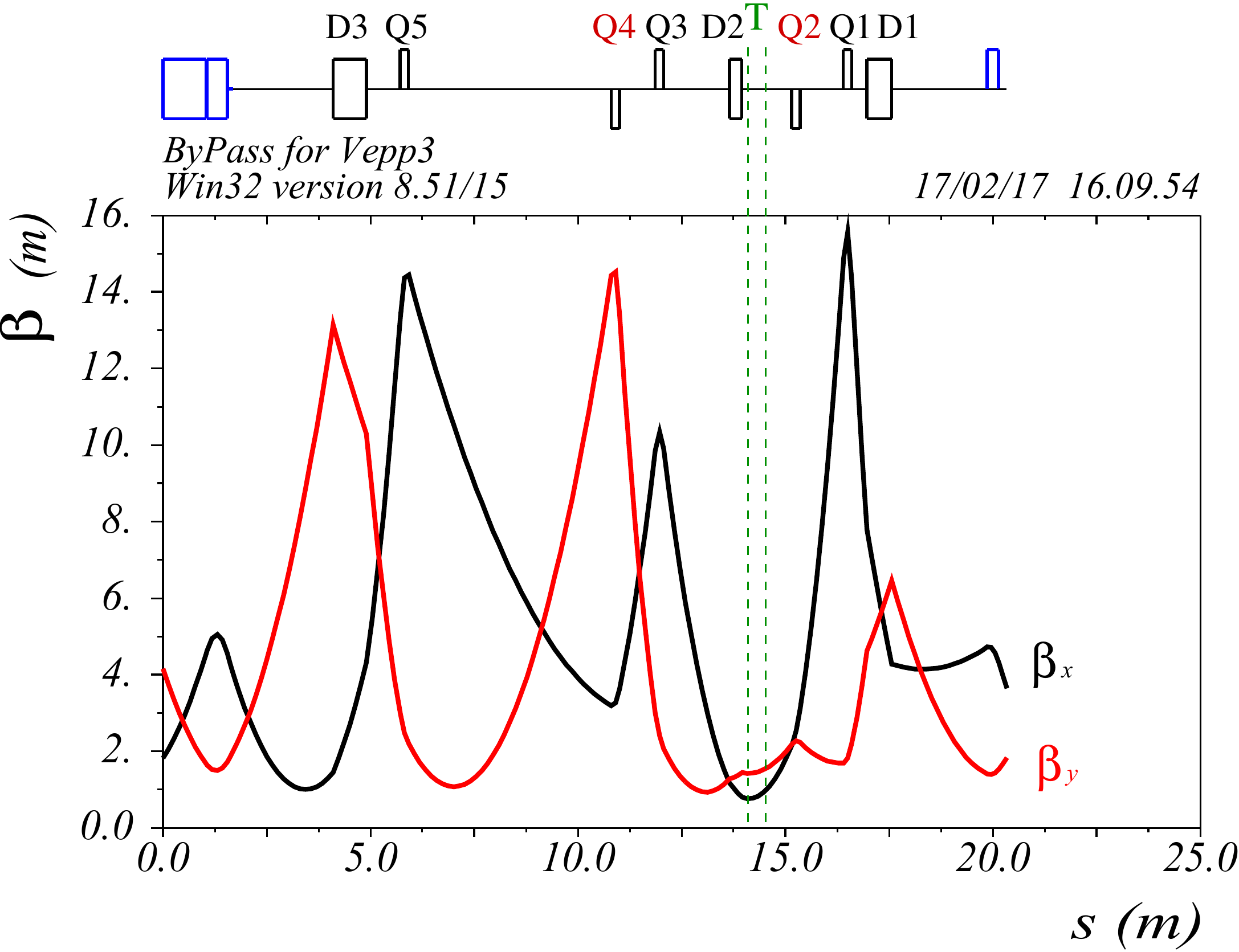}}
\caption[]{\label{fig:betaF}
Calculated beta functions (X and Y) along the bypass for the proposed electron optics. 
Vertical lines show the location of the internal target.
}
\end{figure}

\subsection{Internal target}
Hydrogen gas, flowing through a thin-walled open-ended storage cell cooled to $25^\circ\text{K}$, will be used as an internal target.
The magnetic structure of the bypass is designed in such a way as to provide small values of beta-functions in the location of the storage cell, Fig.~\ref{fig:betaF}, allowing the use of a small--opening cell.
Together with cell cooling this permits us to obtain the required target thickness of about $10^{16}$~atoms/cm$^2$ with a smaller amount of hydrogen gas injected into the target. The gas, leaking out of the cell ends into the ring vacuum chamber, must be pumped out promptly. A set of powerful turbomolecular and cryogenic pumps will be installed in the target chamber,  as well as upstream and downstream from the target chamber.

\subsection{Photon detector}
\label{calorimeter}
The photon detector can be placed at a distance of up to 8~m from the target.
The requirements for the detector are:
\begin{itemize}
\setlength{\itemindent}{-1ex}
\setlength{\itemsep}{-1ex}
\item Energy resolution on the level of $\sigma_E/E=5$\% for photons with energy 
$E_\gamma = 100 - 450$~MeV.
\item Angular resolution on the level of $0.1^\circ$.
\item Angular acceptance as defined by the requirement to detect both photons 
from two-photon annihilation:
\begin{itemize}
\setlength{\itemindent}{-1ex}
\setlength{\itemsep}{-1ex}
\item in $\theta$~: symmetrical range in $\theta_\gamma^{CM}$ around $90^\circ$, e.g.
$\theta_\gamma^{CM}=60^\circ-120^\circ$, which corresponds to 
$\theta_\gamma^{LAB}=1.5^\circ-4.5^\circ$. 
\item in $\phi$~: total 2$\pi$;
\end{itemize}
\item The ability of the detector to sustain a modest photon rate at a level of 1 MHz over its whole area. 
\end{itemize} 

At BINP we have neither an existing calorimeter with suitable parameters nor components to assemble it; 
therefore, we should get it from somewhere else. 
We are going to get crystals from Cornell University, thanks to the collaborators~\cite{Cornell}.

 The end-cap electromagnetic calorimeter of the CLEO-II detector \cite{ref:cleo}
consists of 1600 CsI(Tl) crystals of $5\times 5\times 30~cm^3$ size ($16.2X_0$).
It was used to measure electron and photon energy in a wide range; therefore, 
a direct measurement of its performance at the photon energy of interest for the proposed
experiment  is available:
$$\delta E/E = 3.8\%  \text{~ and~ } \delta x = 12 \text{~mm} \ \  \text{   for } E_\gamma=180\text{~MeV}$$
To ensure a desired angular resolution the CsI(Tl)--calorimeter must be placed
as far as possible from the target, i.e. about 8~m. In this case it would take
about 630 crystals to cover the  required angular range.
A few notes on this detector option should be mentioned:\\
~-- The CLEO-II calorimeter assembly is clearly inappropriate for the proposed experiment,
so a mechanical support must be designed and constructed.\\
~-- CsI(Tl) crystal has a long light emitting time. 
Therefore, its ability to work at a high background rate is limited. 
However, due to the high segmentation of the calorimeter, a long 
output pulse does not seem to be a problem. 
Even for crystals covering the lowest polar angle, the expected rate of background photons is estimated 
to be at a level of a few tens of kHz for the projected luminosity of $10^{33}$ cm$^{-2}$s$^{-1}$. 

The calorimeter should be equipped with front-end electronics and DAQ suitable 
for use in the proposed experiment.

The experiment will require a careful account of the detector responses.
The energy response will be calibrated by using $\gamma\gamma$
coincidence events produced with the  hydrogen target.  
These data will also provide a detector line shape determination.  
The use of the electron beam instead of the positron beam
provides a way to obtain the ``white'' photon spectra without 
the \AB\ signal and the two-photon line. 

\begin{figure}
\includegraphics[width=0.49\textwidth]{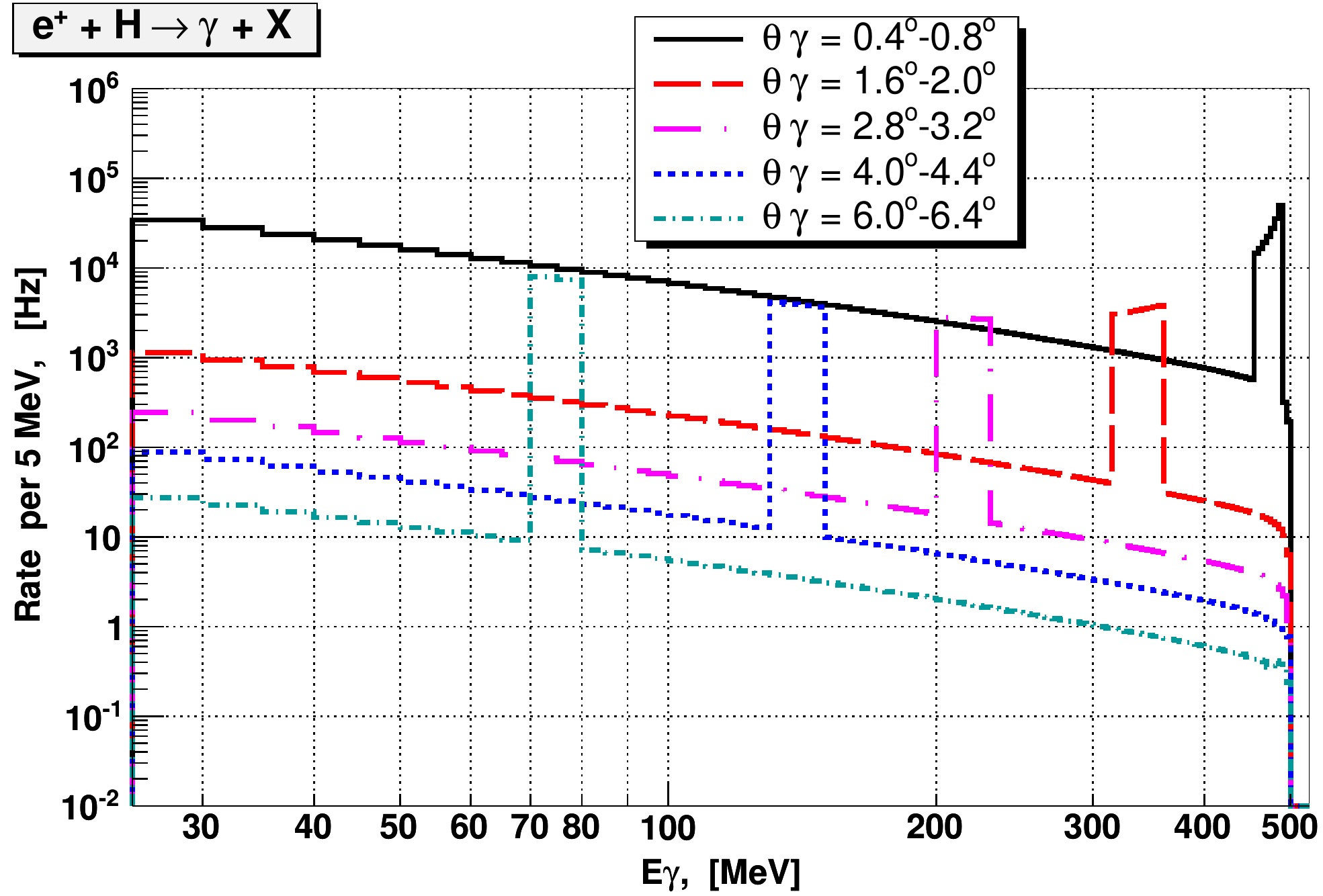}
\hfill
\includegraphics[width=0.49\textwidth]{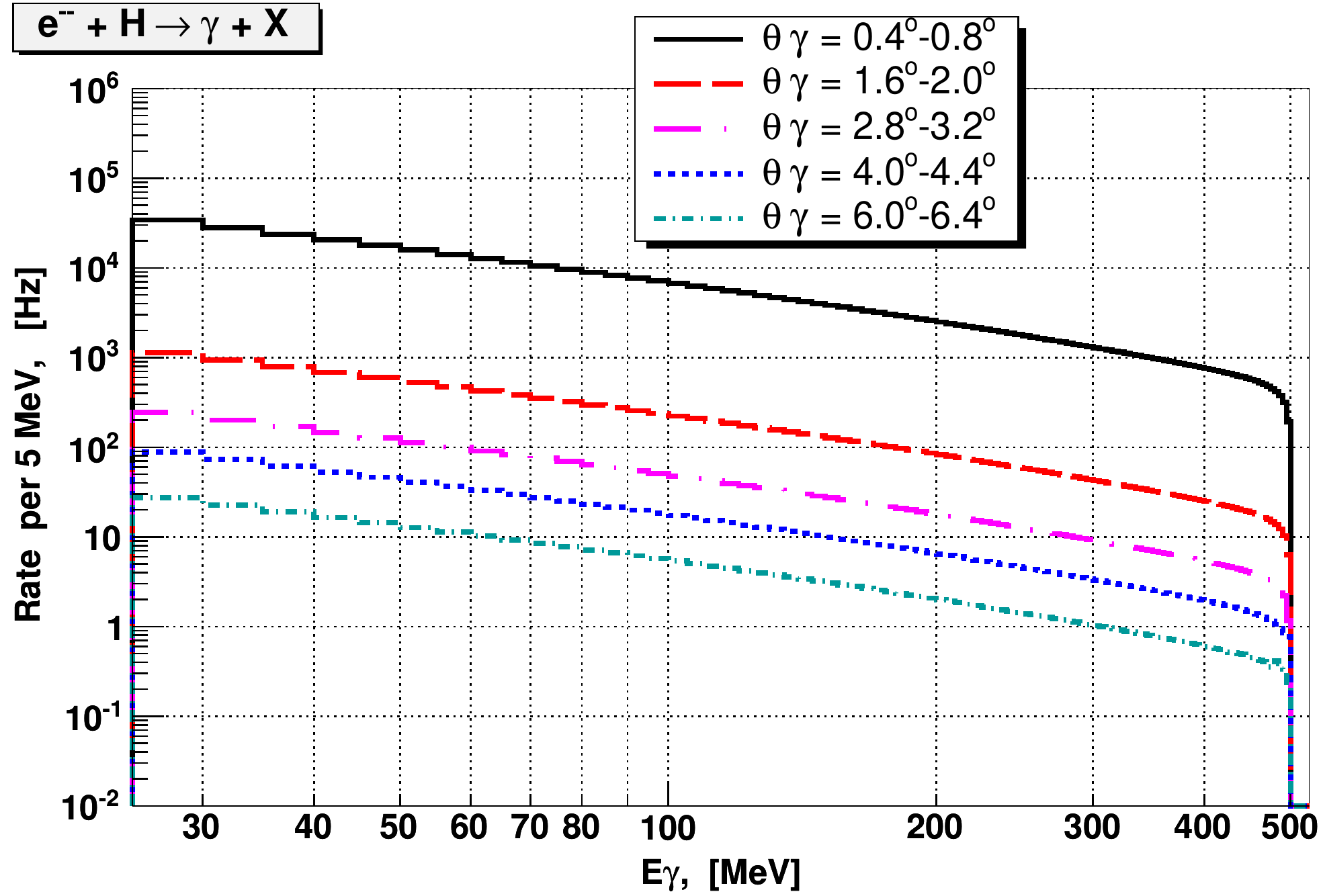}
\caption[]{\label{fig:both}
The photon spectra in the case of a positron beam
(left panel) and an electron beam (right panel) 
incident on an internal hydrogen target with
$10^{16}$~atoms/cm$^2$ thickness.
Beam energy is 500~MeV and beam current is 15~mA in both cases.
Bumps on the left panel, whose positions move with the scattering angle,  
are due to the positron-electron annihilation process.
}
\end{figure}

The left panel of Fig.~\ref{fig:both} shows the result of a calculation of 
the photon spectra for the case of an internal hydrogen gas 
target of $10^{16}$~atoms/cm$^2$ thickness ($5\cdot10^{-10}X_0$), 
a positron beam current of 15~mA (i.e. a luminosity of $10^{33}$~cm$^{-2}$s$^{-1}$) and a 500~MeV 
positron beam energy.
For a narrow range of scattering angle the photon spectrum is composed of a wide background of bremsstrahlung 
process and a peak of the two-photon annihilation process, which moves with the scattering angle.
In the case of an electron beam (Fig.~\ref{fig:both}, right panel), 
the energy spectrum is smooth; this will be used for the calibration 
of the detector response.

\subsection{Positron veto counter}

\begin{figure}
\parbox[c]{0.3\linewidth}{
  \includegraphics[width=\linewidth]{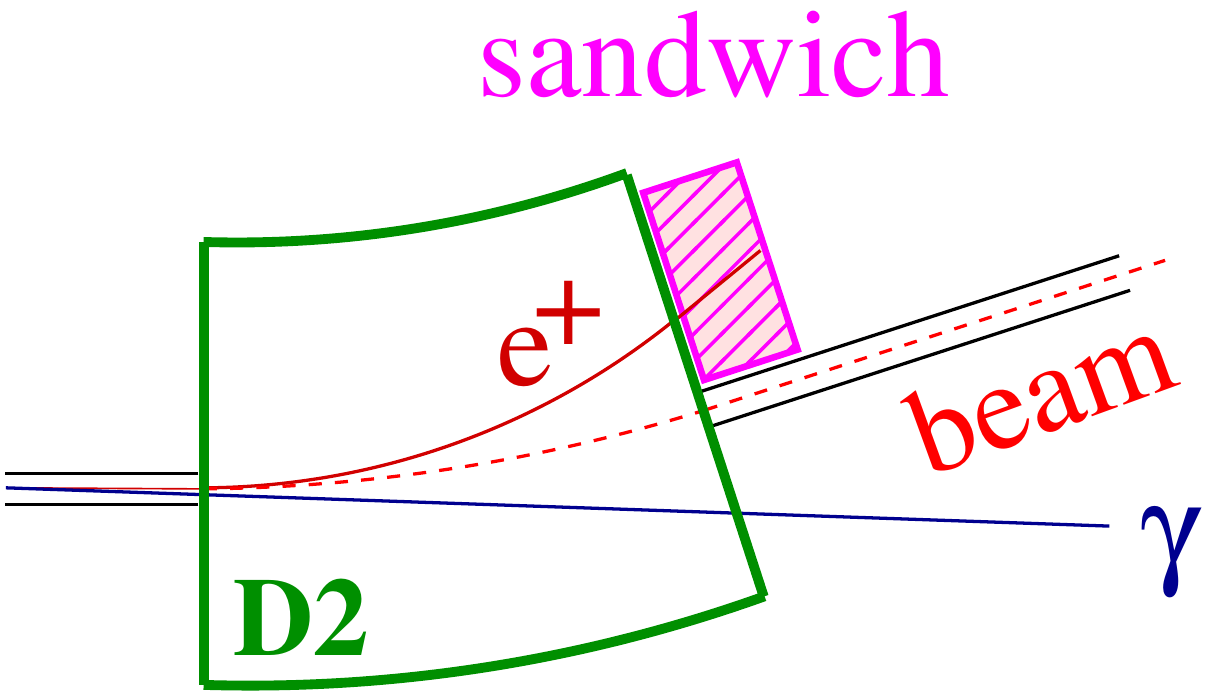}
}
\hfill
\parbox[c]{0.6\linewidth}{
  \includegraphics[width=\linewidth]{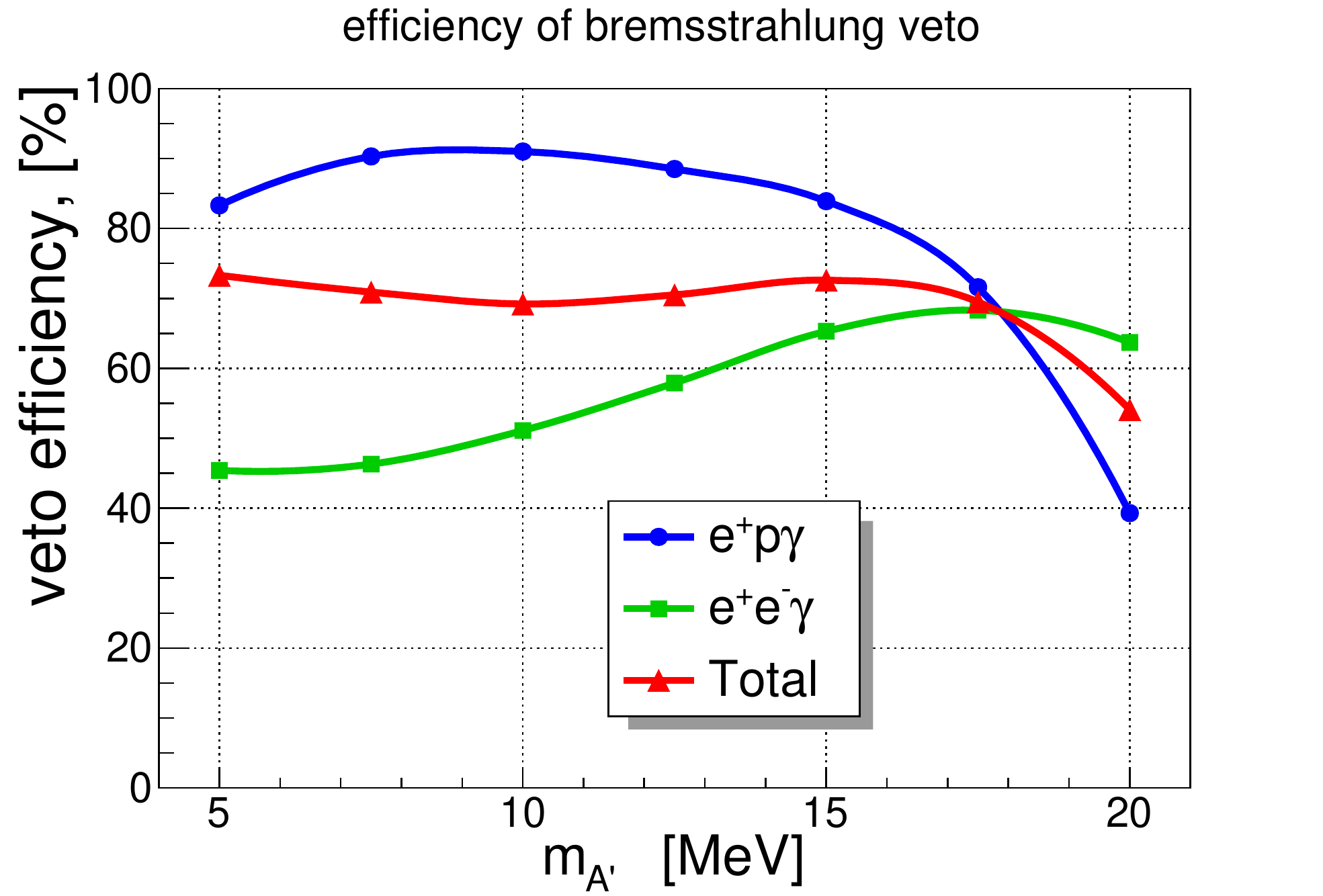}
}
\caption[]{\label{fig:veto}
Schematic layout of the positron veto detector placed behind the D2 dipole magnet, and its efficiency as a function of missing mass.}
\end{figure}

The main single--photon QED background comes from positron bremsstrahlung on hydrogen.
Therefore, a rejection of bremsstrahlung events with an efficiency $\epsilon$ would result in an increase 
in the search sensitivity by a factor of $1/\sqrt{1-\epsilon}$.
Since in this process the positron loses energy and is swept out by the D2 dipole magnet, such background events can be vetoed by detecting the scattered positron.
For this purpose, compact sandwich or simple plastic counters will be installed downstream from the D2 magnet. 
The fraction of bremsstrahlung events detected by these counters depends on their geometry. 
The total rate of positrons emitting a photon with energy above the $E_\gamma=25$~MeV threshold 
can be estimated by integrating Eq.~\ref{eq:tsai}.  
For the luminosity of $L=10^{33}$~cm$^{-2}$s$^{-1}$ the rate is 30 MHz, see Fig.~\ref{fig:rates}. 
Hence,  the veto counter would be useless at this luminosity if it detects all positrons including those which emit photons outside the photon detector acceptance.
Using the MC simulation we have chosen a configuration with a positron rate of $f_{e^+}\sim 2.5$~MHz, 
which gives only 10\% event loss due to accidental coincidences -- see Fig.~\ref{fig:veto}. 
The simulation has shown that the veto efficiency is a function of missing mass range and it is generally higher for 
the bremsstrahlung from a hydrogen target process than for the radiative Bhabha one, with an overall mean value of about 70\%.

\begin{figure}
\parbox[c]{0.3\linewidth}{
\includegraphics[width=\linewidth]{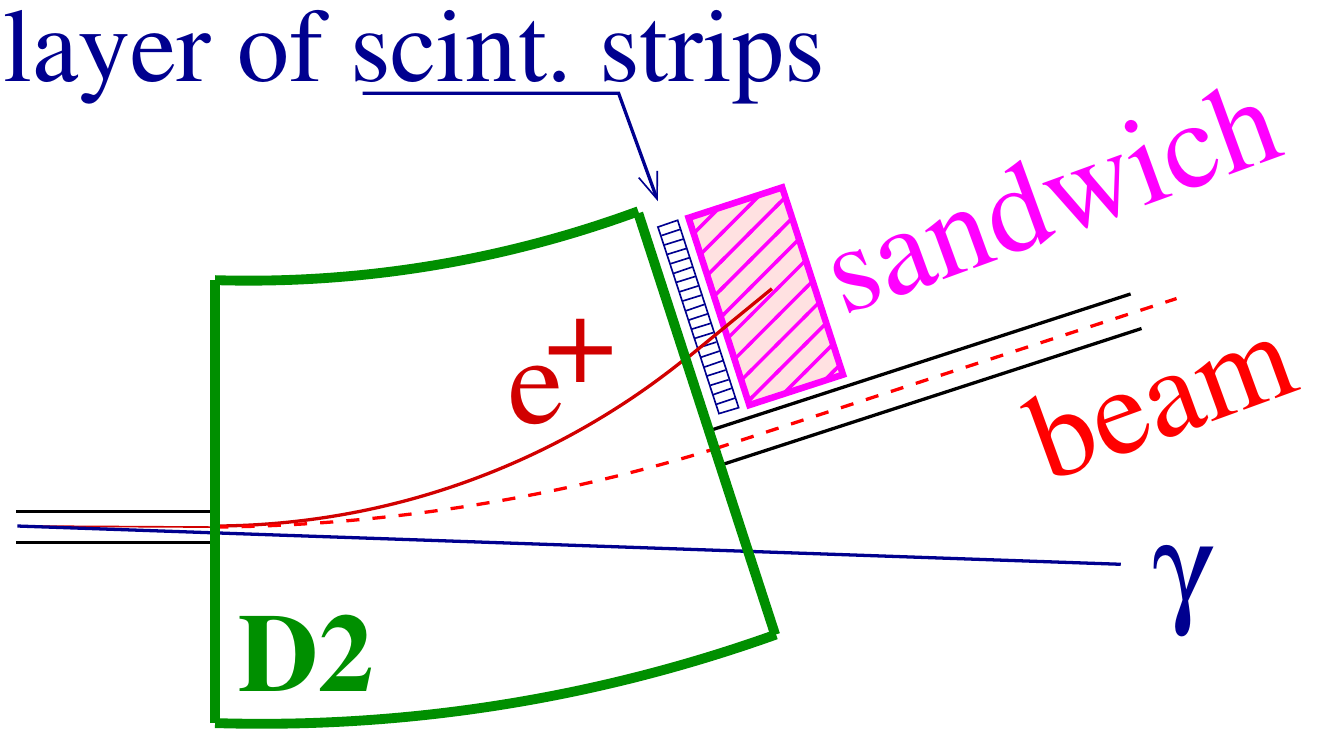}
}
\hfill
\parbox[c]{0.6\linewidth}{
  \includegraphics[width=\linewidth]{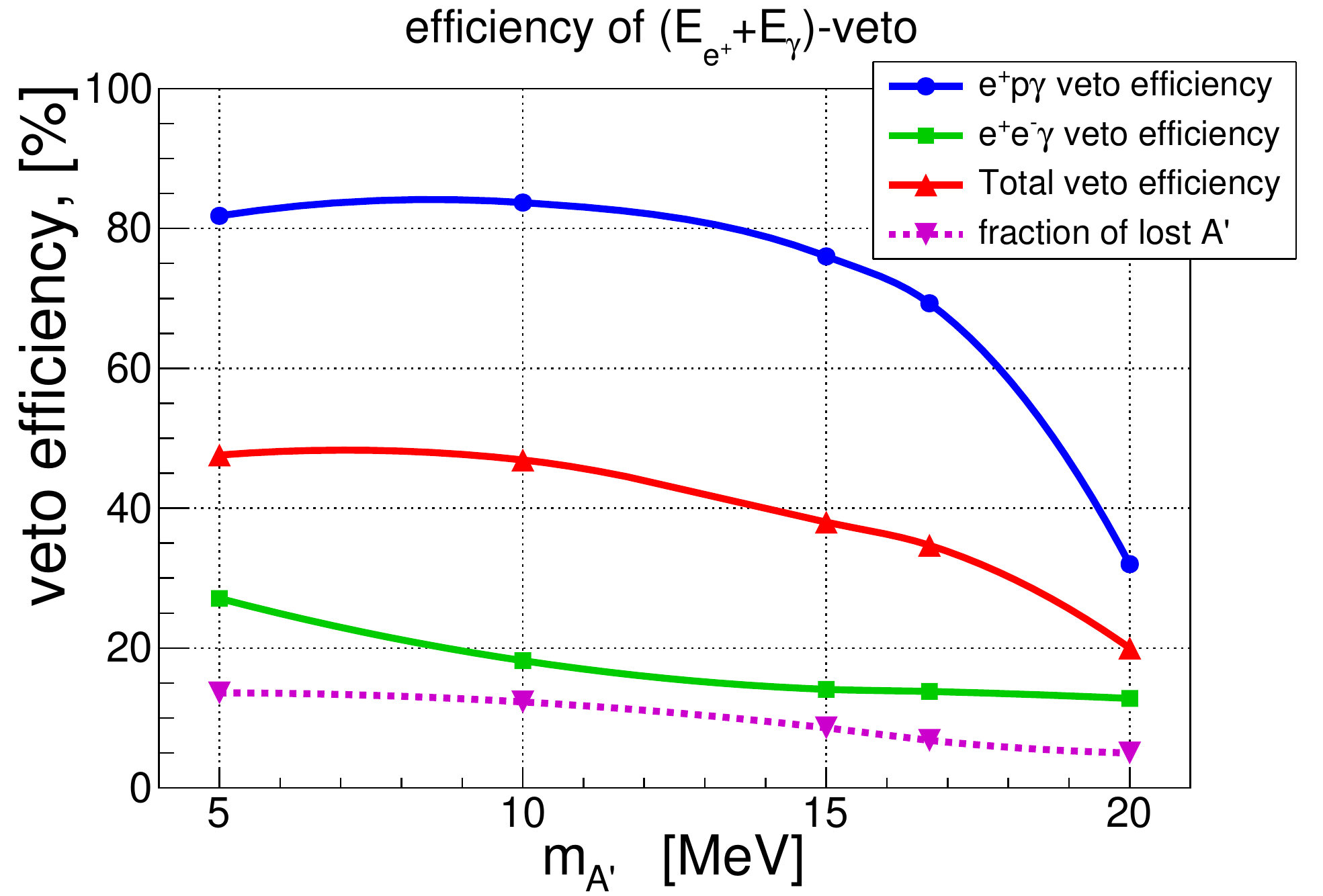}
}
\caption[]{\label{fig:vetoX}
Positron veto detector with a layer of scintillation strips, capable of reconstructing positron energy, and its efficiency as a function of missing mass.}
\end{figure}

However, in the case of a visible decay of $\AP$ to an electron/positron pair, the use of such a veto counter results in a loss of signal events 
because positrons from the $\AB$~decay often hit the veto counter.
We have considered a simple extension -- a layer of scintillation strips placed in front of the veto sandwich, Fig.~\ref{fig:vetoX}. 
This allows us to reconstruct the positron energy and, together with the photon energy reconstructed in the calorimeter, allows 
us to distinguish between the $\AP$ decay process and the bremsstrahlung on the proton. 
However, the veto efficiency for radiative Bhabha events is rather low in this case. 
Nevertheless, such a relatively inexpensive configuration of the veto counter does improve the search sensitivity 
and we are going to implement it.

\subsection{Background from QED annihilation process}
Conventional QED $e^+e^-$ annihilation processes have two or more gamma quanta in the final state,
while in the signal process there is strictly one photon. This can be used to reject such background.

\begin{figure}
\includegraphics[width=0.49\linewidth]{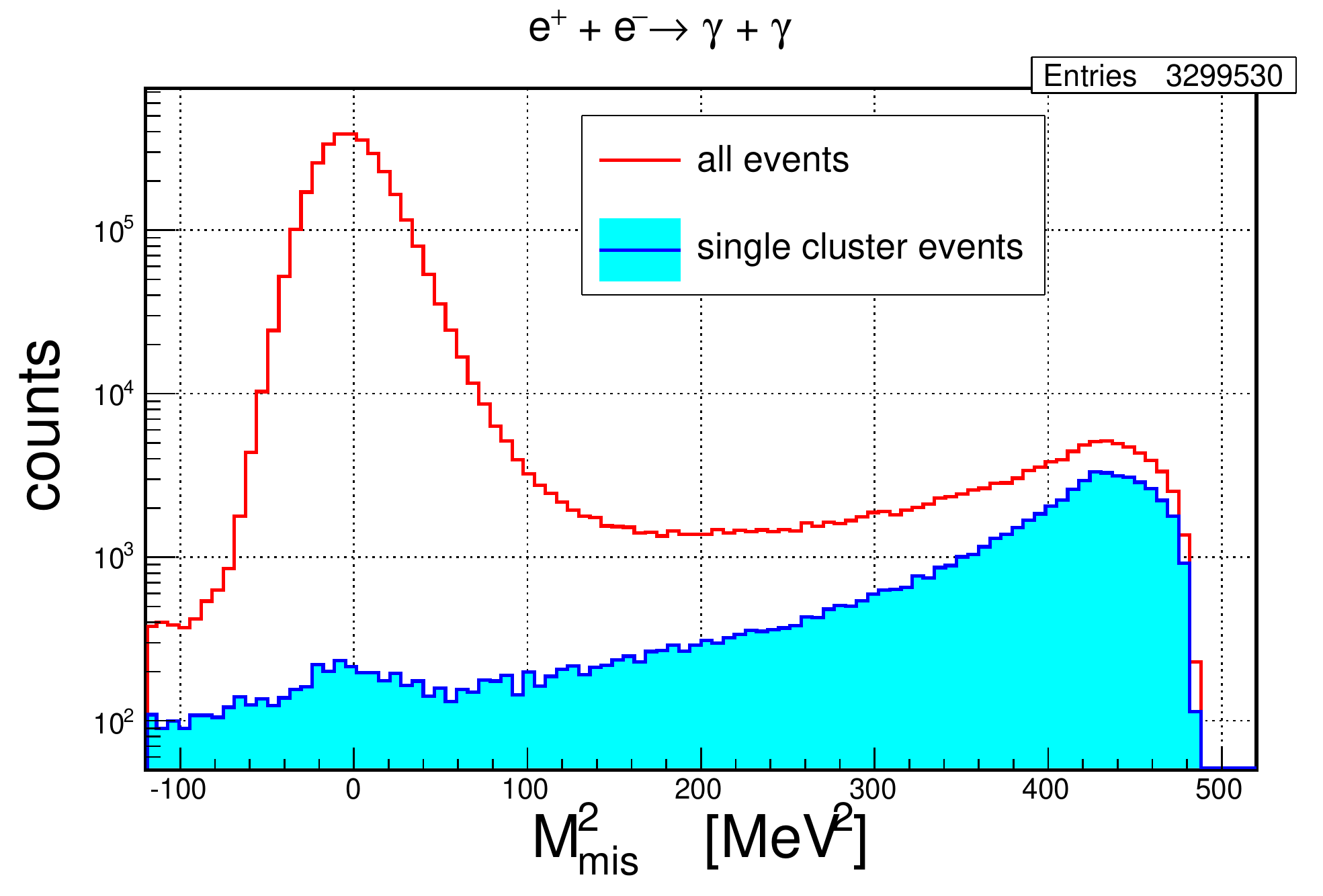} \hfill
\includegraphics[width=0.49\linewidth]{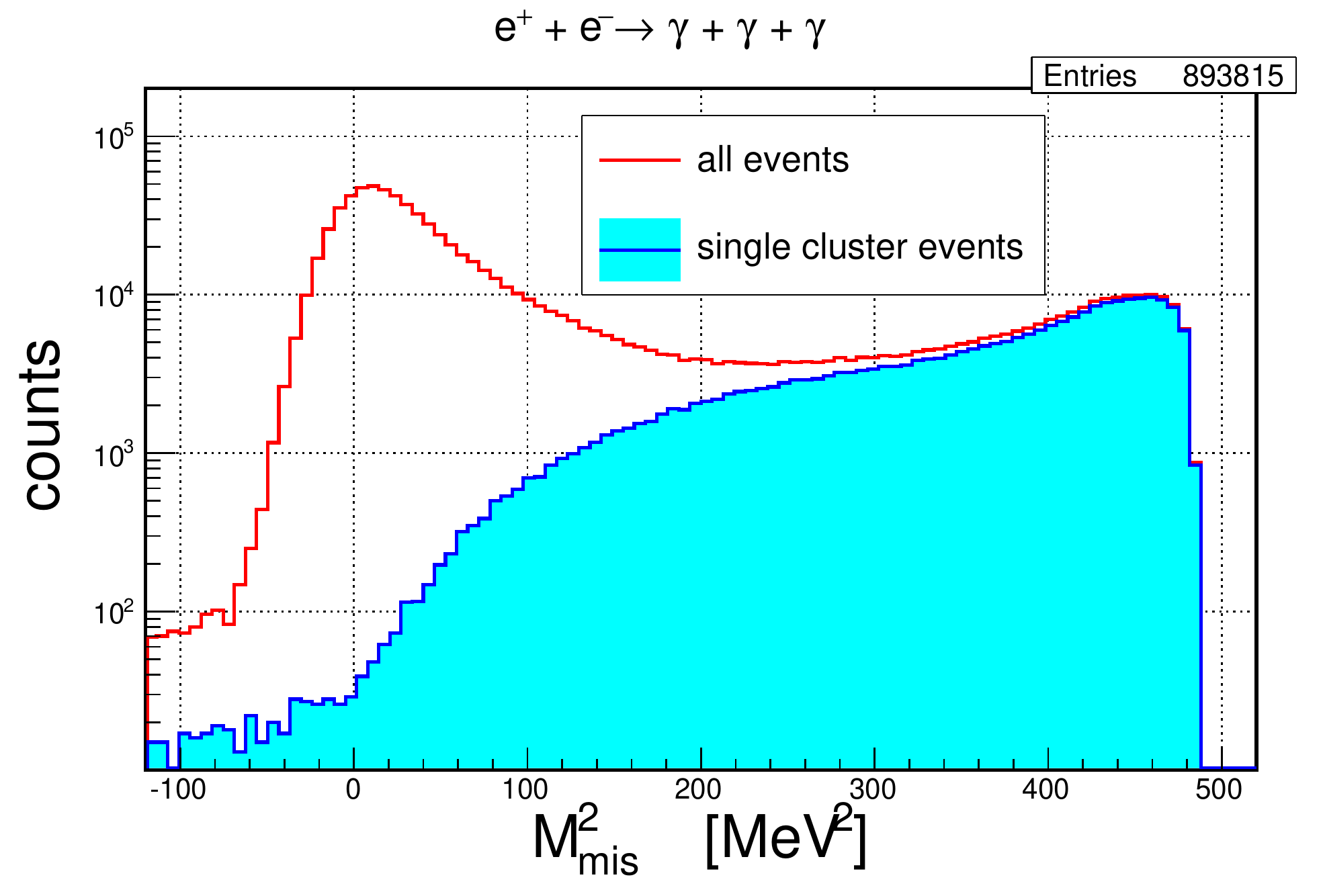} 
\caption{
 \label{fig:ngamma}
 Monte Carlo simulation of the missing mass spectra for the events of 2-$\gamma$ and 3-$\gamma$ background processes, detected by the calorimeter. Requiring a single cluster in the calorimeter one can effectively reject such backgrounds for a low missing mass region.
}
\end{figure}

For this purpose the acceptance of the photon calorimeter is chosen to be symmetrical with 
respect to $\theta_\gamma^{cm}=90^\circ$ in the center of mass frame of an electron/positron pair. 
This means that for the 2-photon annihilation, the calorimeter will always detect either both photons or neither. 
Therefore, the 2-photon annihilation events will be rejected very effectively.
Three-photons events will also be largely rejected, which is demonstrated in Fig.~\ref{fig:ngamma}. This selection cut is especially effective at lower missing mass region, e.g., for $|M^2_{mis}|<100$~MeV$^2$ the suppression factors are 630 and 130 for 2-$\gamma$ and 3-$\gamma$ background processes, respectively.

\subsection{Secondary background sources}
Secondary sources of background events should also be considered. 
Such events are produced when high energy particles (photons, positrons, electrons) hit the materials inside and outside the target area. 
Special efforts should be devoted to installing shielding wherever it is applicable. 
Also, since most secondary showers, born outside the target area, contain electrons and positrons, an additional suppression of such background 
can be achieved by the installation of charge veto counters.

\begin{figure}
\begin{center}
\fbox{\includegraphics[width=0.75\textwidth]{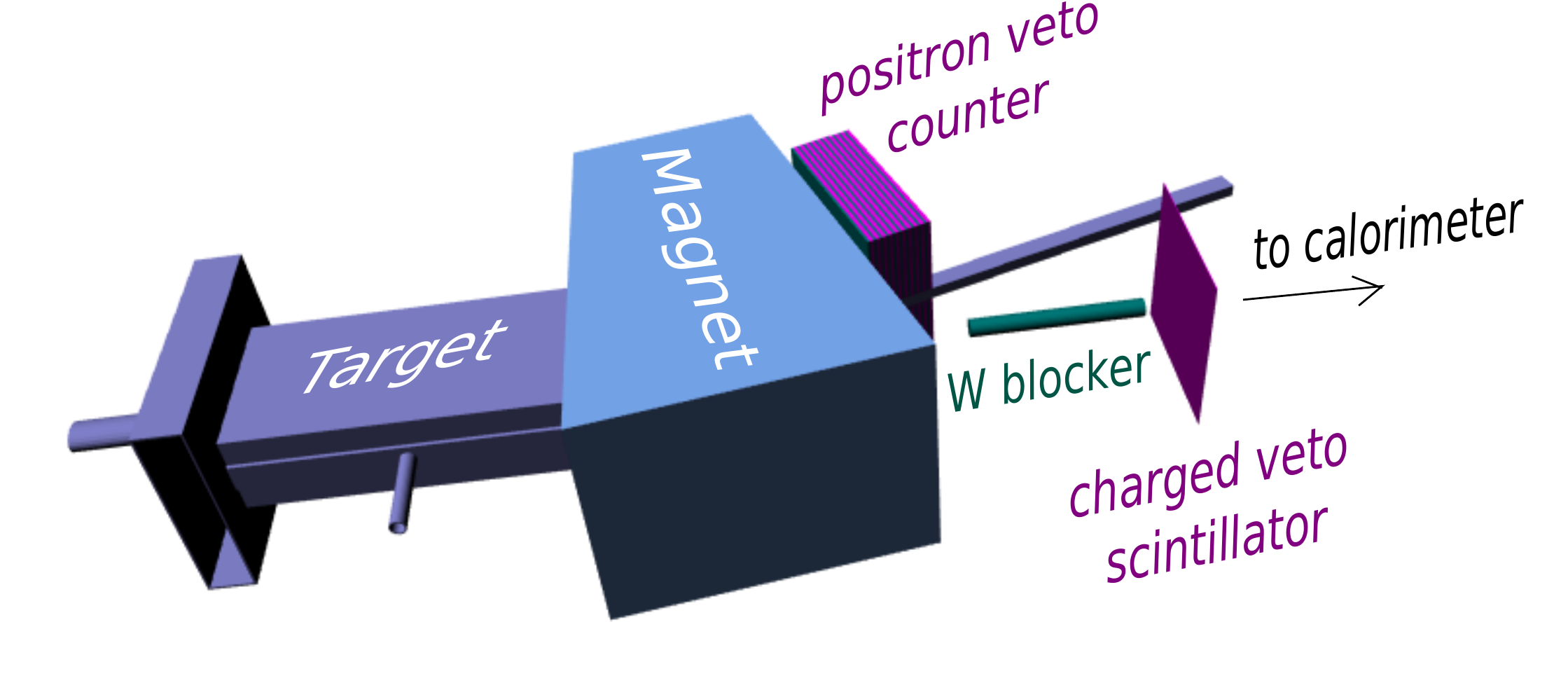}} 
\vskip+\baselineskip
\includegraphics[width=0.6\textwidth]{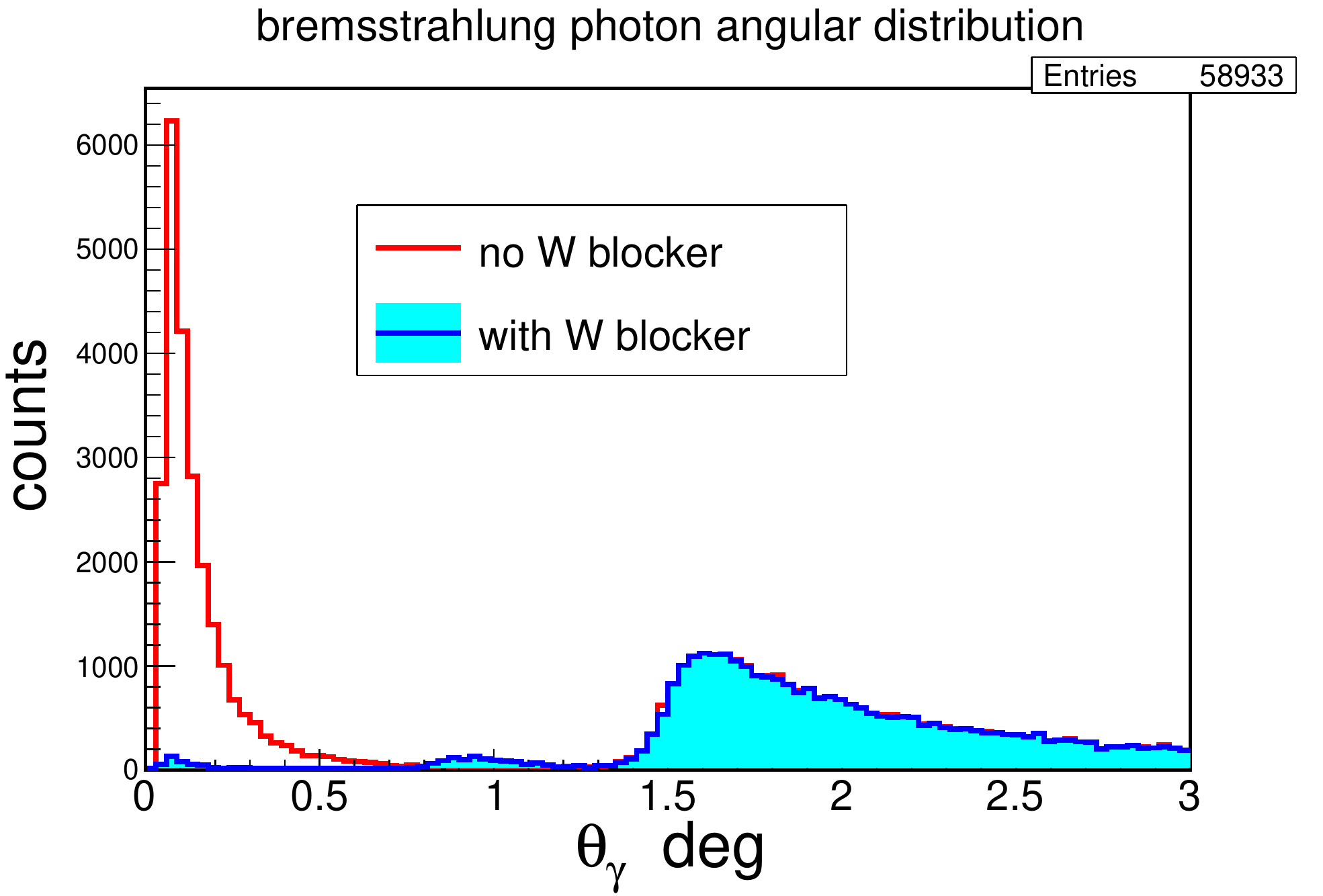}
\end{center}
\caption{
\label{fig:blocker}
The ``blocker'' (tungsten rode of 20~cm in length, 1~cm in diameter) and charged veto scintillator, covering the calorimeter acceptance, installed downstream from the target. The histogram shows the Monte Carlo simulation of the angular distribution of those bremsstrahlung photons which produced a large signal in the calorimeter. The blocker stops forward emitted photons, thus removing the corresponding background.
}
\end{figure}

In the proposed configuration we consider the installation of a tungsten rod (a ``blocker'') to dump a high flux of photons, emitted from the target 
at small angles outside the acceptance of the photon detector, and a thin scintillator, covering the calorimeter acceptance, to veto charged  particles. 
Both elements are installed close to the exit of the dipole magnet, see Fig.~\ref{fig:blocker}.

\subsection{Data Acquisition}
The expected  rate of events with a threshold of minimum energy deposition in the calorimeter $E_\text{cal}>25$~MeV is about 500 kHz for the projected luminosity of the experiment.
The front-end and digitizing electronics should be based on Flash ADC and FPGA logic to provide on-line timing, 
cluster-finding and zero-suppression. 
Similar or substantially faster systems are now widely used or being designed for Data Acquisition in various experiments.
Assuming a conservative value for the on-line suppression factor of 3 for combined veto-channels and a factor 10 for events with two or more photons in the calorimeter, one obtains a 150 kHz final trigger rate. 
This gives an easily manageable data rate of about 50 MBit/s.

\section{Monte Carlo simulation}
\label{montecarlo}
\begin{figure}
\centerline{\includegraphics[width=0.9\textwidth]{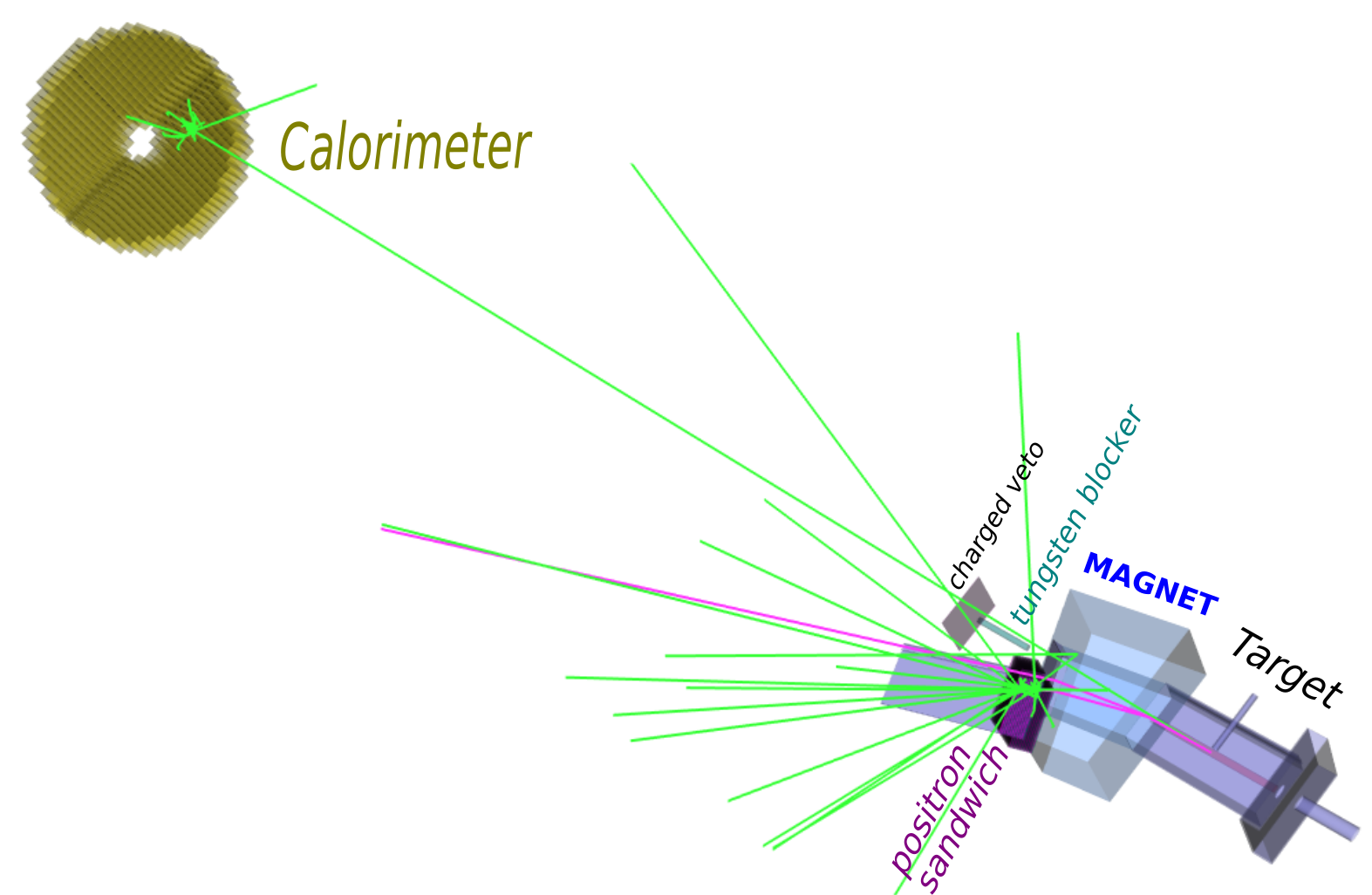}}
\caption{\label{g4model}GEANT4 model of the target area and the photon detector.}
\end{figure}

Thorough Monte Carlo simulation of the measurement procedure, detector response and data analysis has 
been done using a GEANT4 toolkit and a set of dedicated event generators. 
A detailed model of the target area and the photon detector has been developed, see Figure~\ref{g4model}. 

The signal process was generated using the cross section formula \ref{csect}.
The following background processes were taken into account:
\begin{itemize}
\setlength{\itemindent}{-1ex}
\setlength{\itemsep}{-1ex}
\item Single photon bremsstrahlung of positron on proton: The expression \ref{eq:tsai} is used;
\item Single photon bremsstrahlung of positron in elastic (Bhabha) scattering on atomic electrons: 
A simplified version of the event generator described in \cite{CMD} is applied;
\item Two- and three-photon annihilation: A procedure outlined in \cite{berends} is adopted for the event simulation.
\end{itemize}

\begin{figure}
\includegraphics[width=0.48\textwidth]{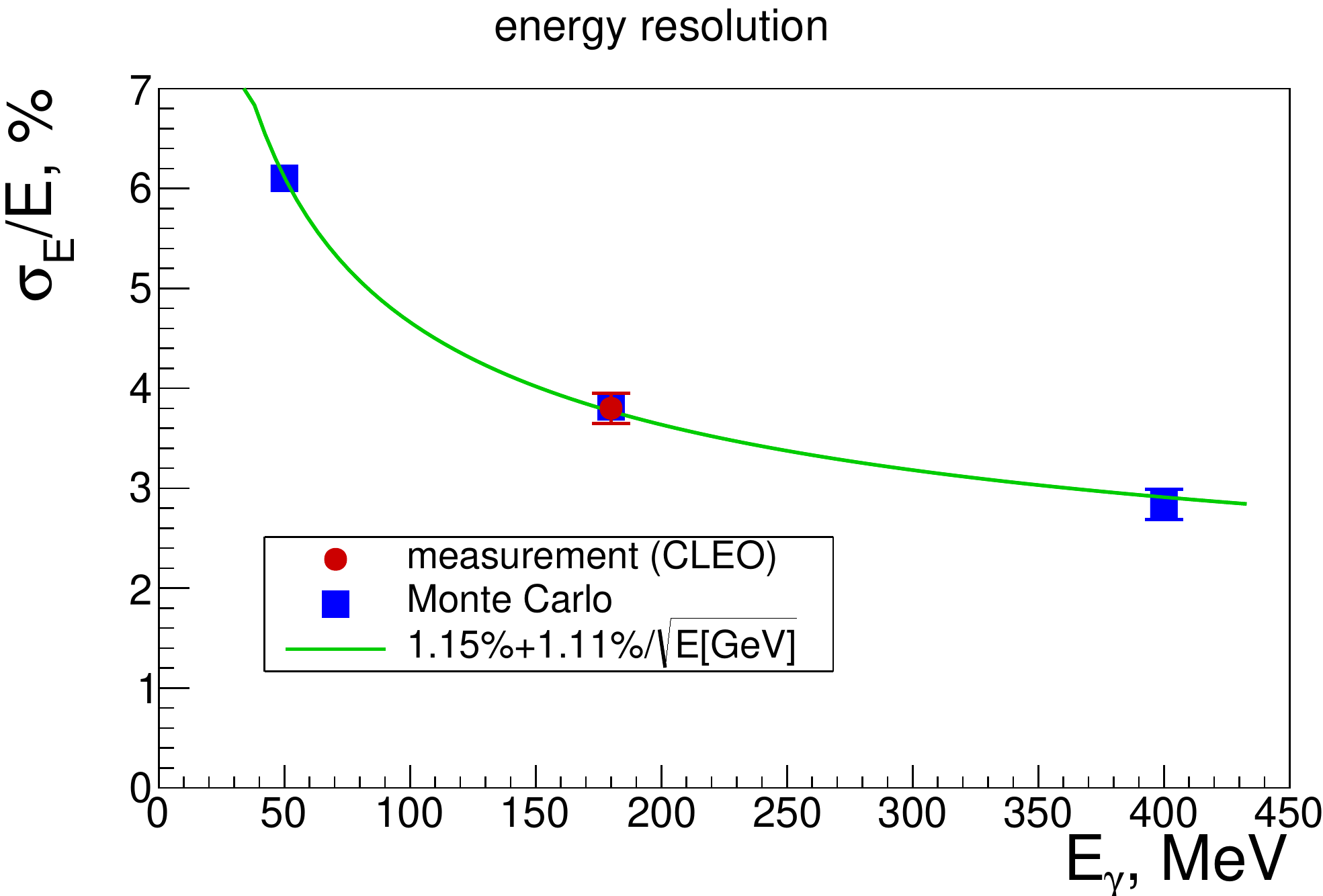}
\includegraphics[width=0.48\textwidth]{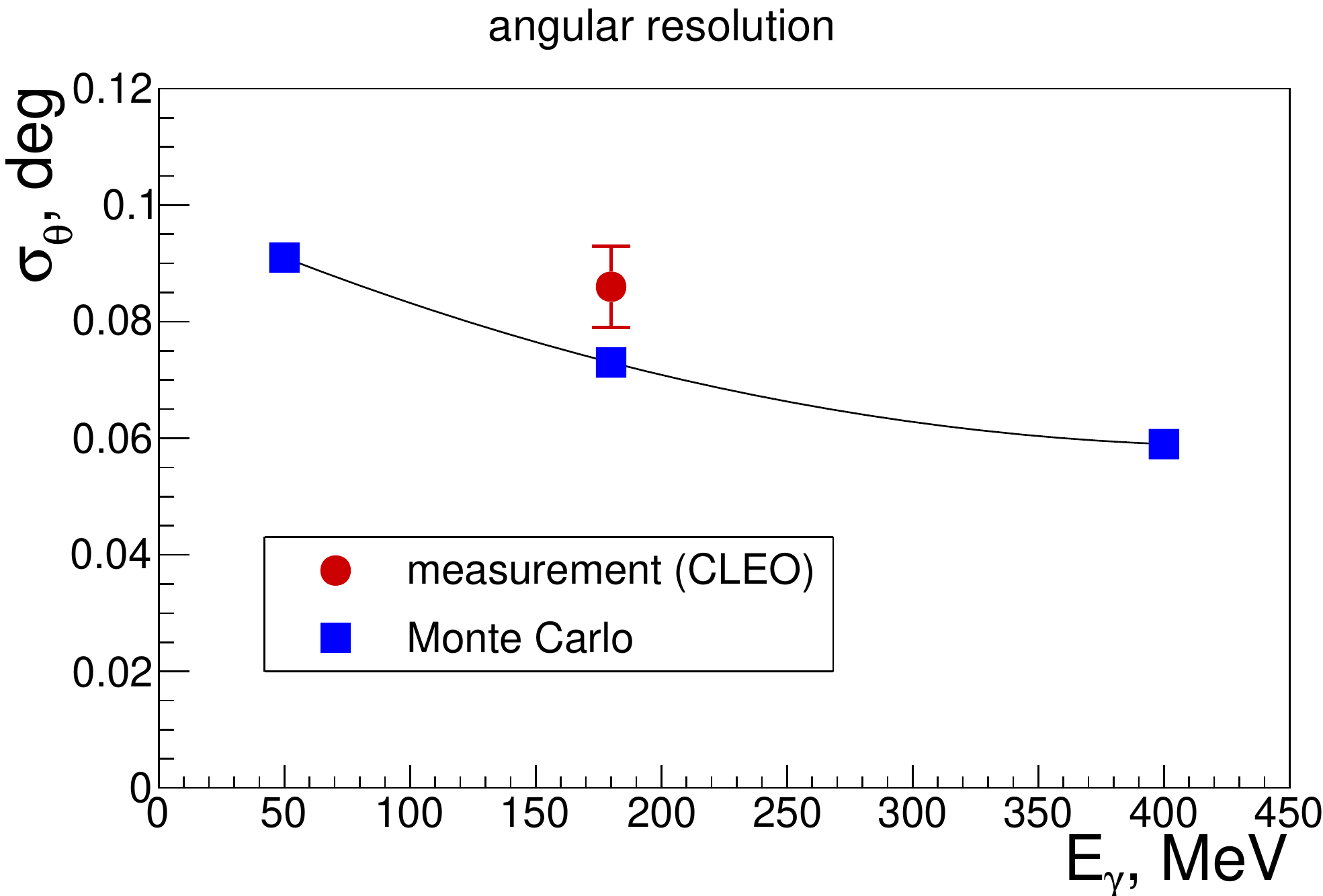}
\caption{
\label{fig:eresol}
Energy (left panel) and angular (right panel) calorimeter resolutions. 
Squares -- from Monte Carlo simulation, circles -- from the measurement \cite{ref:cleo}.
}
\end{figure}

We have compared the calorimeter resolutions from simulation and those reported in the CLEO calorimeter paper \cite{ref:cleo}. 
It was found that if we add the noise of electronics with parameters taken from \cite{ref:cleo} to the fluctuation of 
energy deposition obtained in GEANT4 simulation, then the measured and the simulated resolutions match nicely, see Fig.~\ref{fig:eresol}.

\begin{figure}
\parbox[c]{0.65\linewidth}{\includegraphics[width=\linewidth]{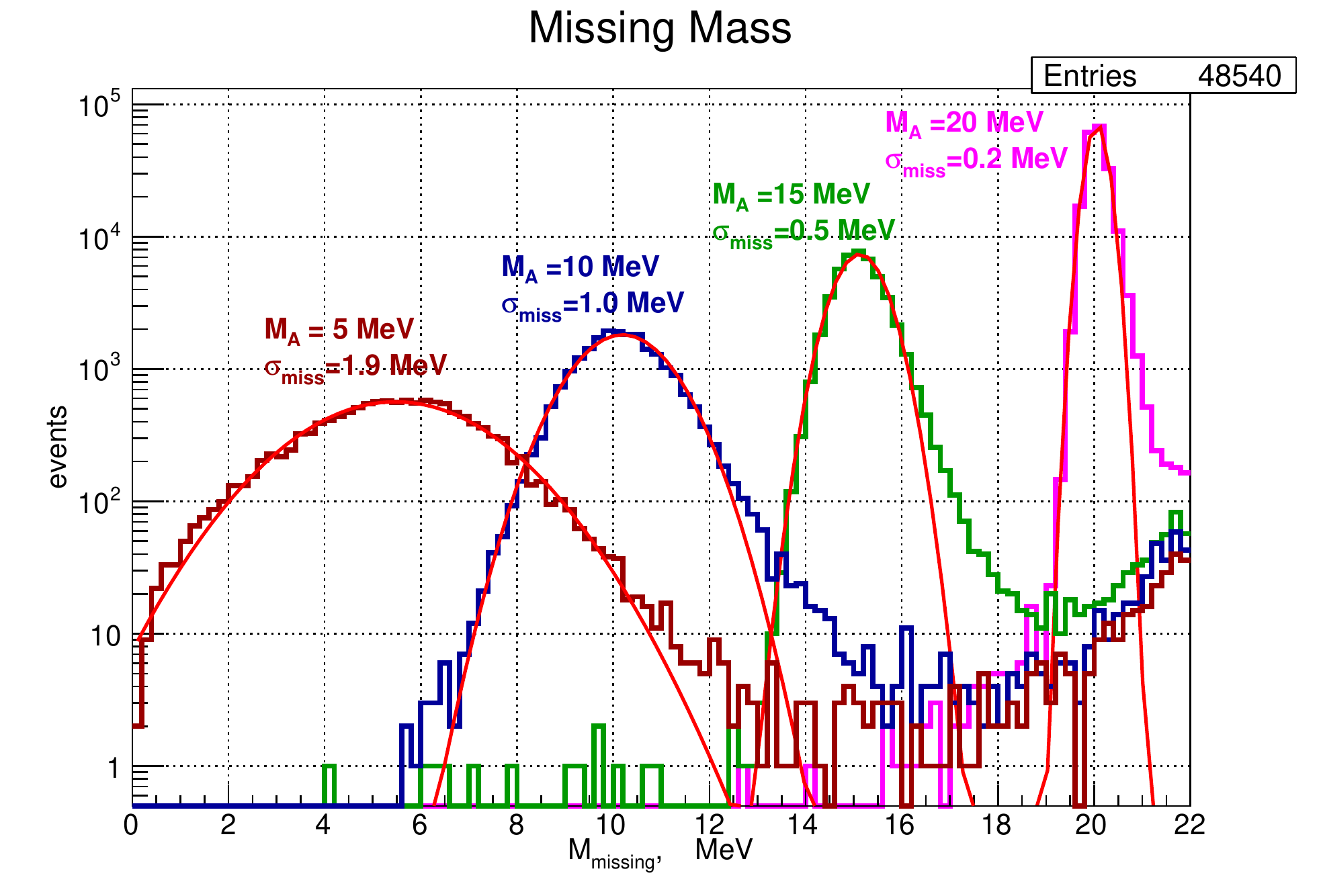}}
\hfill
\parbox[c]{0.35\linewidth}{\includegraphics[width=\linewidth]{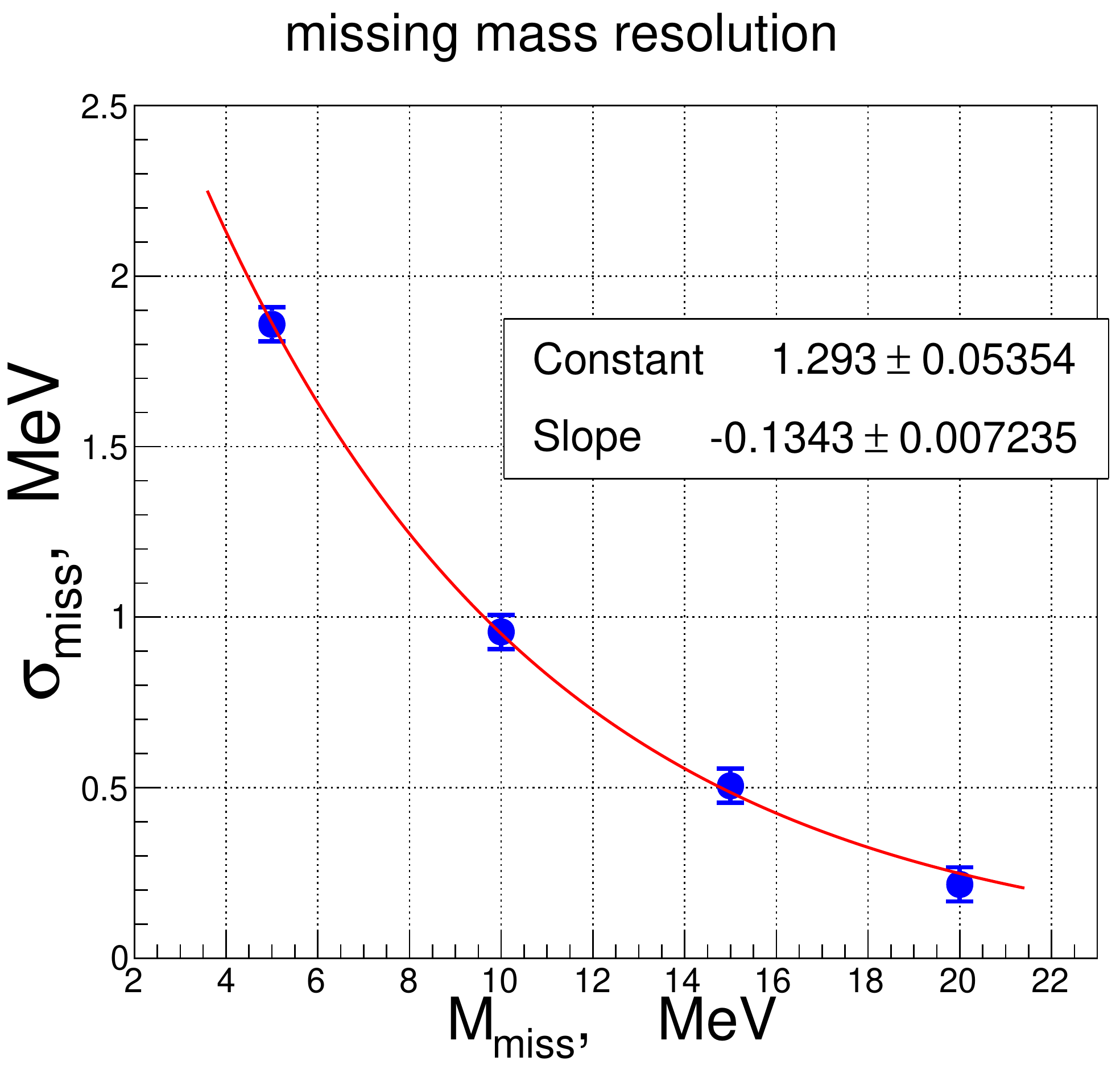}}
\caption{
\label{fig:mresol}
Monte Carlo simulation of the missing mass resolution for four values of the missing mass: 5, 10, 15 and 20 MeV.
}
\end{figure}

The task of data analysis is to select events with a single cluster in the calorimeter and no signal in the veto counters 
and to search  for a peak in the missing mass distribution in selected events,
so we need to know the experimental resolution for a reconstructed missing mass. 
Figure~\ref{fig:mresol} shows the results for four values of \AB-boson masses when purely \AB~production events were generated. 
One can see that the resolution improves rapidly with the increase in the mass of the \AB. 
The dependence can be fitted with an exponential function 
\begin{equation}
\label{sigmis}
\sigma_{miss}(M_{A^\prime})=3.6\cdot e^{-0.13\cdot M_{A^\prime}} \left[\text{MeV}\right].
\end{equation}
Note that such behavior means that a big tail from two-photon annihilation ($M_{A^\prime}=0$)
is expected in the missing mass spectrum, which indeed is observed. 
Therefore, for an efficient search at low $M_{A^\prime}$, a strong suppression of 2-photon events will be required. 
That is why the acceptance of the proposed photon detector is chosen to provide the detection of both gamma quanta to improve the search sensitivity at low $M_{A^\prime}$.

\begin{figure}
\includegraphics[width=0.49\linewidth]{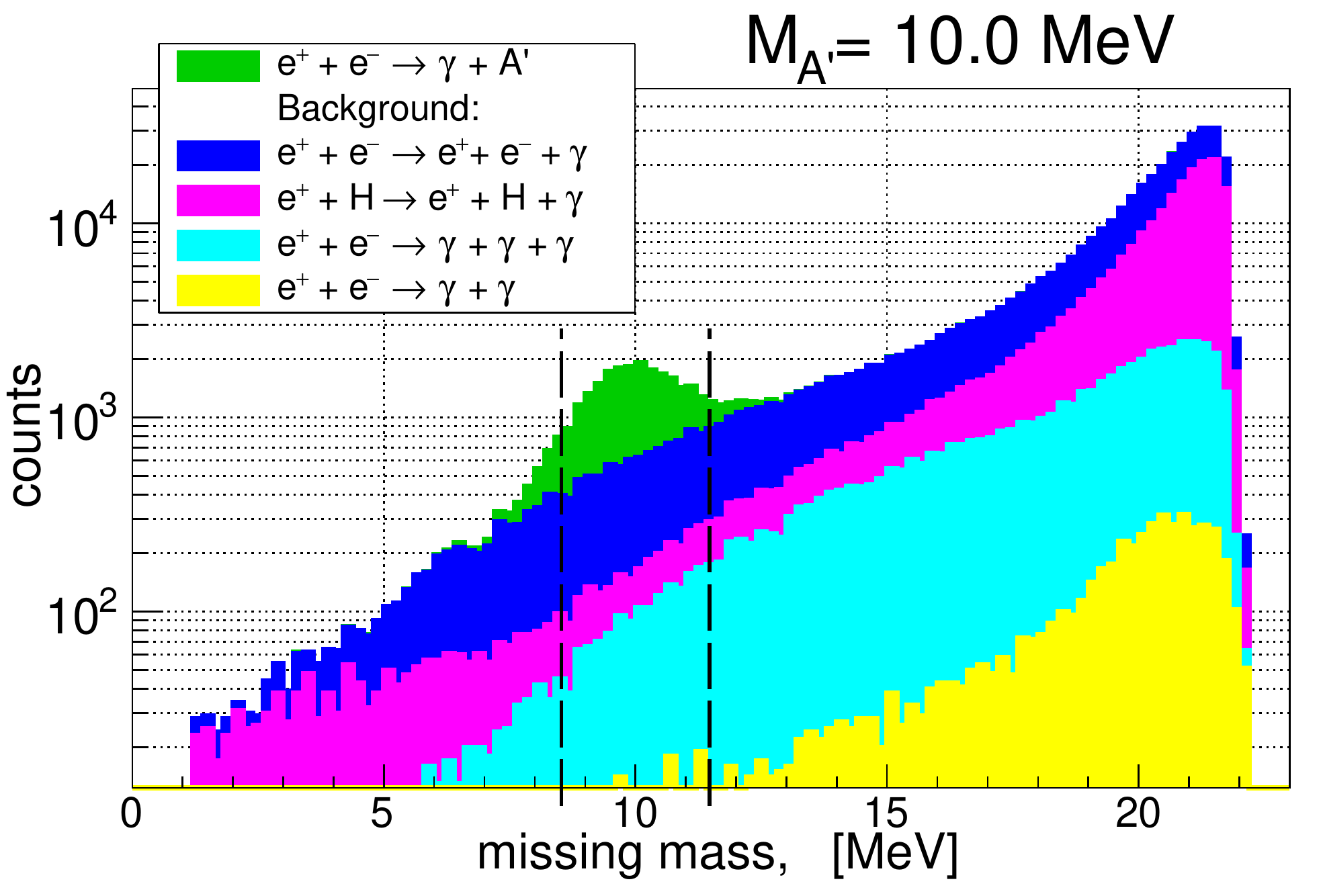} \hfill
\includegraphics[width=0.49\linewidth]{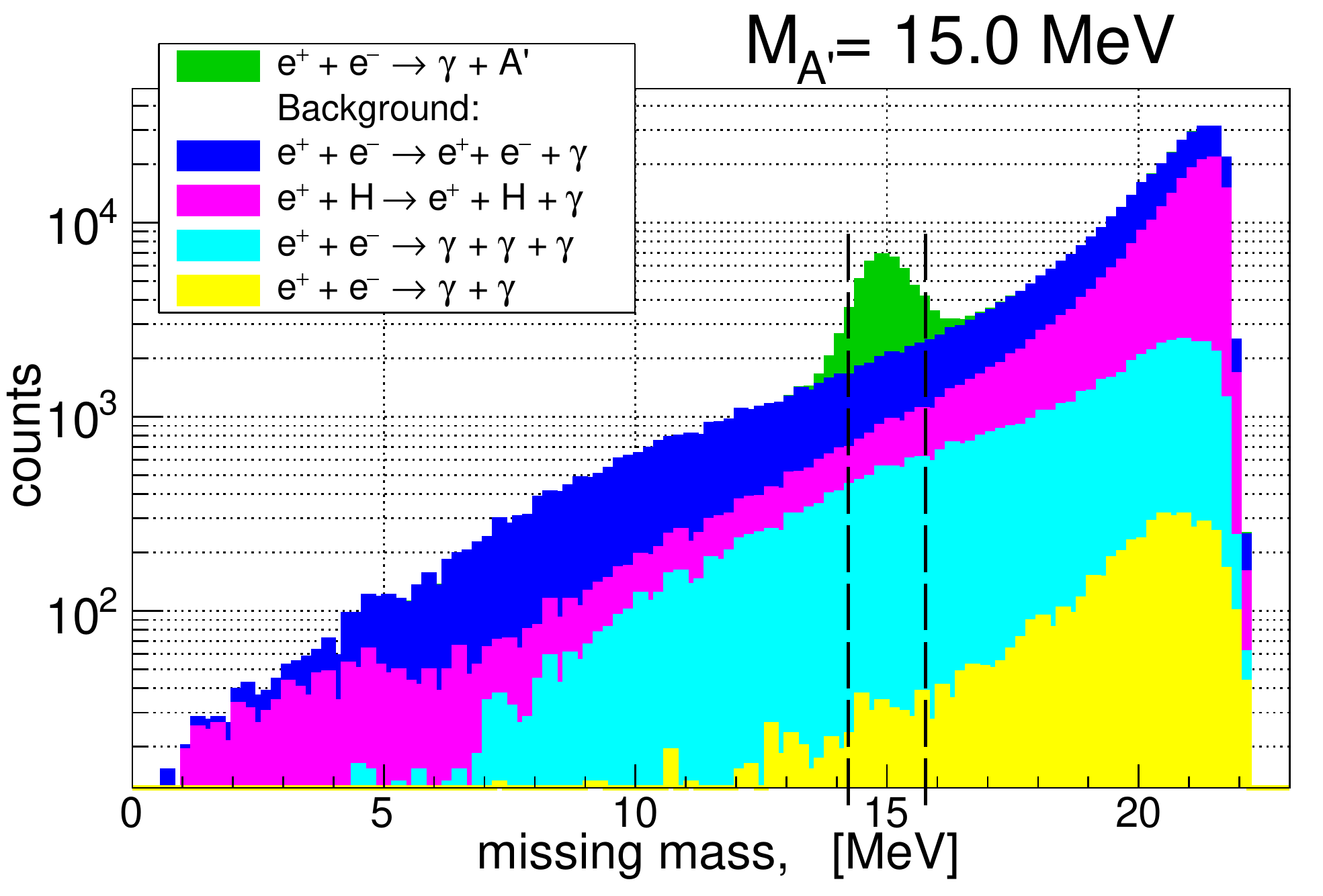} 
\caption{
 \label{fig:mcresults}
 Results of the full Monte Carlo simulation for two possible masses of the \AB~after all selection cuts are applied. 
 Vertical lines indicate the width of a sliding search window.
Stacked histograms show contributions of various background processes to the final missing mass spectrum.
 The mixing constant is taken to be $\epsilon^2=10^{-2}$. 
 For a luminosity of $10^{33}$~cm$^{-2}$s$^{-1}$ such a spectrum can be obtained in 3 seconds.
}
\end{figure}

Figure \ref{fig:mcresults} shows the missing mass spectra obtained in the full Monte Carlo simulation.

\section{Run time and the projected sensitivity}
This simulation data allows estimation of the search sensitivity. 

The search conditions are: 
\begin{itemize}
\setlength{\itemsep}{-1ex}
\item 
Beam energy of 500 MeV;
\item
Luminosity of $10^{33}$ cm$^{-2}$s$^{-1}$, which corresponds, for example, to a beam current 
of 30 mA and a target thickness of $5\times10^{15}$atoms/cm$^2$;
\item
Run time of $10^7$ seconds, which is a half-year run with 65\% time utilization; 
\item
The search is performed using a sliding missing mass window with a width 
of $\pm 2 \sigma_{miss}$, where $\sigma_{miss}$ is evaluated using Eq.(\ref{sigmis}).
\end{itemize}

\begin{figure}
  \centerline{\includegraphics[width=0.7\textwidth]{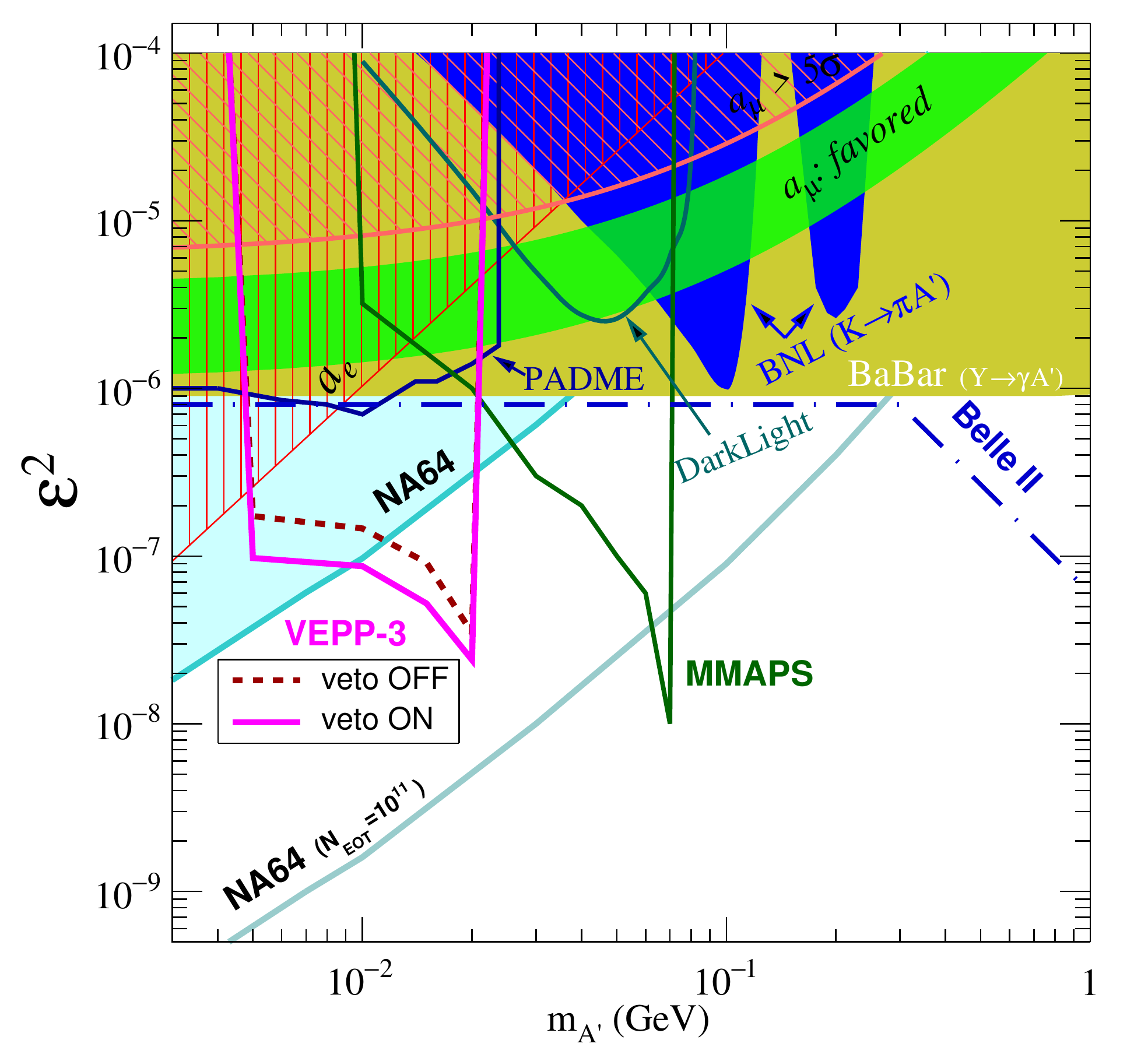}}
\caption[]{\label{fig:result-invis}
Existing and projected upper limits of a coupling constant $\epsilon^2$ of a new boson vs. its mass   
for the case of its invisible decay to dark matter particles.
The hatched areas are regions excluded by the results of the measurements of anomalous
magnetic moments of electron and muon~\cite{po09}. 
The green band indicates a ``welcome'' area, where the consistency of theoretical
and experimental values of $a_\mu$ would improve to $2\sigma$ or less \cite{po09}. 
The shaded areas are the excluded parameter spaces of the completed searches \cite{babar17,bnl1,bnl2,na64}.
Curves show areas of search of the experiment at VEPP-3 and of other proposed experiments \cite{DSW16}. 
The search sensitivity of the VEPP-3 experiment is presented for two operation regimes: with and without use of the positron veto counter.
The veto-OFF search is a factor of two less sensitive, but this is a truly decay-mode independent approach.
}
\end{figure}

\begin{figure}
  \centerline{\includegraphics[width=0.8\textwidth]{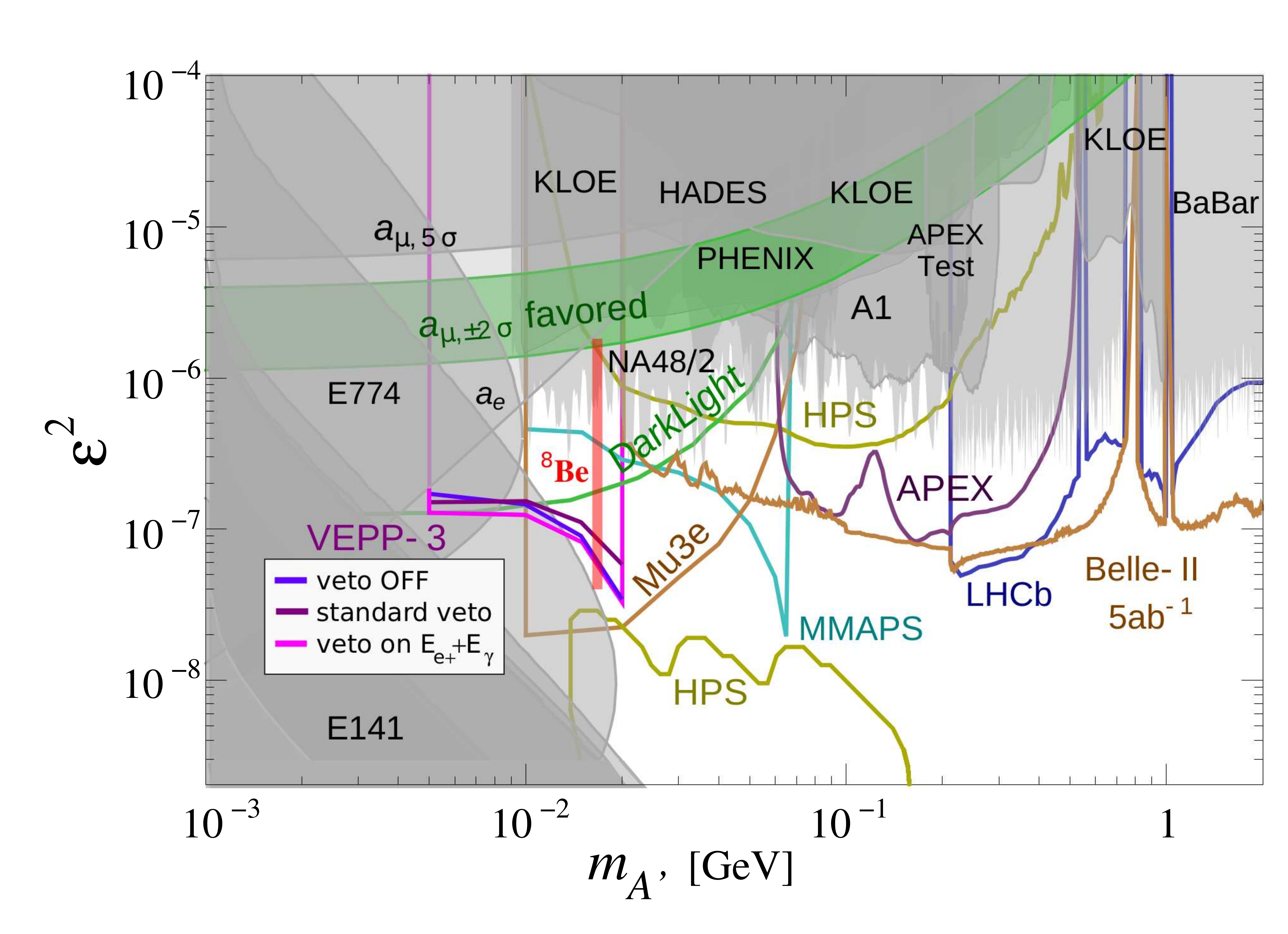}}
\caption[]{\label{fig:result-vis}
Existing and projected upper limits of a coupling constant $\epsilon^2$ of a new boson vs. 
its mass for the case if its decay to $e^+e^-$ or $\mu^+\mu^-$ pairs with a 100\% branching ratio. 
The shaded areas are the results of the completed direct searches: 
beam dump experiments \cite{bd1, bd2}; $e^+e^-$ colliding beam experiments \cite{babar,kloe};
fixed--target experiments \cite{mainz,apres}.
Curves show areas of search of other proposed experiments, see the review~\cite{DSW16}. 
The search sensitivity of the \V3 experiment is presented for various modes of operation of the positron veto. 
Note that the \V3 search area covers almost the whole parameter space allowed for the putative ``Atomki boson''~\cite{atomki}, 
which is proposed to explain an anomaly in $^8$Be$^*$ decay \cite{8be}.
}
\end{figure}

The results of the simulation are presented in Figures~\ref{fig:result-invis}, \ref{fig:result-vis} for a Confidence Level of 95\%. 
Plots of the mixing constant $\epsilon^2$ versus  the mass of the new boson $\mb$ are shown.
The region  which will be accessible for the search in the proposed experiment is outlined,
together with some other completed and proposed measurements.
The \AB~mass range is 5-20 MeV and the reach for $\epsilon^2$ is between $1.7\cdot 10^{-7}$ and $2.4\cdot 10^{-8}$.

\section{Summary}
We propose a sensitive search of an exotic \AB, using a missing mass
reconstruction in a positron-electron annihilation, utilizing the \V3 internal target
facility and the VEPP-5 positron/electron injection complex at
the Budker Institute of Nuclear Physics, Novosibirsk, Russia. 

The key features of the proposed measurement are:
\begin{itemize}
\setlength{\itemsep}{0pt}
\item The missing mass method. No assumptions about decay modes of the \AB\ are required;
\item The mass range for the proposed search is 5-20 MeV, which is not accessible in
most other proposed fixed-target or colliding-beam approaches;
\item Moderate experimental luminosity ($\sim10^{33}$cm$^{-2}$s$^{-1}$) and high segmentation of 
the photon detector, which allows the use of available CsI(Tl) crystals;
\item The use of a veto-detector for scattered positrons and a symmetric angular acceptance of the
photon detector ($\theta^{\text{\tiny{CM}}}_\gamma=90^\circ\pm 30^\circ$), which permits an effective
suppression of the QED background, resulting in an increase in the search sensitivity.   
\end{itemize} 

\noindent
The projected sensitivity for the square of the coupling constant of 
the \AB~to the electron is $\epsilon^2=5\cdot 10^{-8}$ at $\mb=15$~MeV at a CL = 95\%.

\section{Acknowledgements}

We thank Rouven Essig for productive discussions and helpful feedback.
This work was supported in part by the US DOE 
and by the Ministry of Education and Science of the Russian Federation.
Jefferson Science Associates, LLC, operates Jefferson Lab for the US
DOE under US DOE Contract No. DE-AC05-060R23177.
Parts of this work related to the study of design and performance of the projected photon calorimeter and the bypass (sections \ref{bypass}, \ref{calorimeter} and \ref{montecarlo}) 
are supported by the Russian Science  Foundation (project No.14-50-00080).

\end{document}